\documentclass[11pt,a4paper]{article}
\usepackage{jheppub}
\usepackage[T1]{fontenc}
\usepackage[utf8]{inputenc}
\usepackage{hyperref}
\usepackage{amsmath}
\usepackage{amssymb}
\usepackage{mathrsfs}
\usepackage{bbm}
\usepackage{tikz}
\usepackage{subfig}
\usepackage{rotating}
\usepackage{multirow}
\usepackage{chngpage}
\usepackage{array}
\usetikzlibrary{matrix,arrows}

\renewcommand{\imath}{\mathrm{i}}
\renewcommand{\iota}{e}
\newcommand{\SO}{\mathrm{SO}}

\newcommand{\SU}{\mathrm{SU}}
\newcommand{\U}{\mathrm{U}}
\newcommand{\E}{\mathrm{E}}
\newcommand{\GL}{\mathrm{GL}}

\numberwithin{equation}{section}

\title{Non--Abelian orbifold compactifications of the heterotic string}

\author[a,b]{Sebastian J.~H.~Konopka}
\affiliation[a]{Physik Department T30, Technische Universit\"at M\"unchen,
James--Franck--Stra\ss e, 85748 Garching, Germany}
\affiliation[b]{Arnold--Sommerfeld--Center for Theoretical Physics, Ludwig-Maximilians-Universit\"at M\"unchen,
Theresienstra\ss e 37, 80333 M\"unchen, Germany}

\emailAdd{sebastian.konopka@physik.uni-muenchen.de}
\subheader{TUM-HEP-860/12\\LMU-ASC 70/12}

\abstract{
I consider the construction of heterotic orbifold models
based on a toroidal orbifold with non--Abelian point group. I construct
an explicit model based on the point group $S_3$ and calculate the
spectrum and remnant symmetries. This model provides a simple example
of rank reduction of the Yang--Mills gauge group directly in the
string theory rather than in the effective field theory. 
}


\arxivnumber{1210.5040}

\begin{document}
\maketitle

\section{Introduction}
String theory is the most advanced framework to provide a unified
quantum description of all fundamental forces in nature. It turned out
that the heterotic string
\cite{HeteroticString1,HeteroticString2,HeteroticString3} has very
good properties for string model building \cite{Raby,Antoniadis1,Aldazabal}. To
arrive at a 
phenomenologically appealing model the number of spacetime dimensions
has to be reduced from ten to four. One way to do so is to compactify
six internal dimensions onto a toroidal orbifold. String kinematics on
orbifold backgrounds have been studied extensively in the literature
ever since its first treatment in
\cite{StringOrbifolds1,StringOrbifolds2}.
Nevertheless does the restriction to $\E_8\times\E_8$ heterotic
orbifold models leave us with a plethora of ways to choose the
background geometry and fluxes. All possible six dimensional toroidal
orbifolds leading to an $N \geq 1$ supersymmetric spectrum have been
recently classified in \cite{FischerOrb}. Most of the orbifold models
constructed up to today use orbifolds with Abelian point groups. By
using appropriate Wilson line configurations it has been possible to
obtain the exact MSSM spectrum using Abelian orbifolds with point
group $\mathbb{Z}_2\times\mathbb{Z}_2$ \cite{Blaszczyk1,Kappl1} or
$\mathbb{Z}_6^{II}$ \cite{Lebedev3}.

In most Abelian constructions the rank of the gauge group is not
reduced. One possibility to achieve rank reduction in the string model
itself is to use a
non--diagonal embedding of the space group into the gauge group
\cite{Ibanez1,Ibanez2}. In the construction of \cite{Ibanez2} the
space group acts on the bosonic coordinates $X^I$ describing
the Yang--Mills sector by affine transformations, $X^I \rightarrow
\theta^I_J X^J + 2\pi \lambda^I$, where $\theta^I_J$ is a lattice
automorphism. Hence, by the non--Abelianess of the space group the
rank of the gauge group is reduced. However, performing explicit
spectrum calculations is quite involved. The main reason is the
distinction of the bosonic formulation between Cartan currents $H^I =
\mathrm{i} \partial X^I$ and root currents $E_\alpha = \exp\left(
  \mathrm{i} \alpha X \right)$, $\alpha$ some root of $\E_8
\times \E_8$. The realization of the Yang--Mills sector by free
holomorphic fermions provides a more symmetric treatment. It is a
hybrid of the free fermion formulation
\cite{Antoniadis1,Dreiner,Antoniadis,Kakushadze} and the usual bosonic
formalism. Instead of using a pure free fermion CFT to describe the
internal sector of the model, only the Yang--Mills sector is described
by a fermion CFT \cite{WittenNA}.

In this paper, I consider an alternative approach to rank reduction
via a non--Abelian point group. I provide an explicit
example that shows that rank reduction is possible at the string
level. The construction is based on the free fermion realization of
the Yang--Mills sector and uses bosonization techniques to analyze the
spectrum. Non--Abelian heterotic orbifold
models have been previously constructed using the free fermion
construction in \cite{Kakushadze,Aoki}. For an analysis of discrete
torsion in such models see \cite{Feng}.

In section \ref{sec:stringorbi}, I
review the construction of a generic orbifold model and
describe the structure of the state space that is essential for
understanding the orbifold projection. The material 
presented here is well known and the purpose of this section is to fix
notation. Section \ref{sec:model} parametrizes the family of models
under consideration. Apart from the orbifold geometry the only additional
parameter is the embedding of the point group in the gauge group. In
the section \ref{sec:projecting}, I deal with the
projection process itself. Here, I describe an algorithm for the
extraction of the transformation properties of string states from the
specification. Then, I construct an explicit example in section
\ref{sec:example}. This model is based on a toroidal orbifold with
point group $S_3$. I demonstrate the algorithm by giving intermediate
results and show that the remnant gauge and discrete symmetries do not
suffer from anomalies in the low energy effective field theory
description. Finally, section \ref{sec:conclude} presents my
conclusions. This paper contains two appendices. In appendix
\ref{app:s3} properties of the group $S_3$ and its representation
theory are reviewed and in appendix \ref{app:fermionicakm} the
fermionic realization of affine Ka\v{c}--Moody (AKM) symmetries is reviewed.

\section{Strings on orbifold backgrounds}
\label{sec:stringorbi}
\subsection{The uncompactified heterotic string}
An orbifold state in heterotic string theory is defined by an orbifold
CFT. The mother CFT is given by a $(0,1)$ supersymmetric
$\sigma$-model on the worldsheet $\Sigma$ where the fields take values
in ten dimensional Minkowski space times a holomorphic $\E_8\times\E_8$
torus \cite{HeteroticString1}. The worldsheet CFT is non--interacting
and the state space $V$ factorizes as
\begin{align}
  V &= V_{\mathrm{ghosts}} \otimes V_{\mathrm{4d}} \otimes V_{\mathrm{bosonic}} \otimes
  V_{\mathrm{fermionic}} \otimes V_{\mathrm{YM}}.
\end{align}
The non--unitary CFT described by $V_{\mathrm{ghost}}$ accounts for the
overcounting due to the invariance under local
superdiffeomorphisms. As all orbifold constructions considered here
are compatible with this local symmetry, I do not consider this
subsector henceforth. The subsector $V_{\mathrm{4d}}$ describes propagation in
the uncompactified directions, while $V_{\mathrm{bosonic}}$ and
$V_{\mathrm{fermionic}}$ do so for the six internal
dimensions. Moreover, forgetting about worldsheet SUSY, I distinguish
between the fermionic and bosonic components of the embedding
superfields $X^\mu$ and $\bar{\psi}^\mu$, $i=0,\ldots,9$. $V_{\mathrm{YM}}$ is
responsible for the $\E_8\times\E_8$ Yang--Mills symmetry from the spacetime point 
of view. There are several ways to realize this subsector. The only
restrictions are that its central charge $c$ be $c = 16$ and this
CFT be purely holomorphic. In the bosonic description $V_{\mathrm{YM}}$ is
given by $16$ holomorphic bosons taking values in a torus
defined by the root lattice of $\E_8\times\E_8$. Alternatively, one
can realize it by $16 + 16$ holomorphic free Majorana--Weyl fermions $\Xi^I(z)$
with GSO projections being different for the first $16$ and second
$16$ fermions. This way, one reproduces the CFT of the bosonic
description \cite{WittenNA}, cf. appendix \ref{app:fermionicakm}.

The spectrum of the mother CFT enjoys two important symmetries. The
first one is Yang--Mills symmetry. It originates from the invariance
of the Yang--Mills fermion CFT under orthogonal transformations within
both of the groups of $16$ fermions $\Xi^I(z)$. The generators
$\Omega^{IJ}(z)$ of the $\SO(16)\times\SO(16)$ symmetry are given by
\begin{align}
  \Omega^{IJ}(z) &= \mathrm{i} :\Xi^I(z)\Xi^J(z):.
  \label{eq:ymcurrents}
\end{align}
The zero modes of these currents are the generators of the gauge
symmetry. Moreover, by adding the zero modes of the spin fields
$S^\alpha(z)$ that change boundary conditions from NS to R the gauge
group is enhanced to $\E_8\times\E_8$ as can be shown by passing to
the bosonized description. 
The second important symmetry is ten dimensional Poincar\'e
symmetry. Under a Poincar\'e transformation $(\Lambda,a)$ the fields
$X^\mu$ and $\bar{\psi}^\mu$ transform as
\begin{subequations}
  \begin{align}
    (\Lambda,a) \cdot X^\mu &= \Lambda^\mu_\nu X^\nu + a^\mu \\
    (\Lambda,a) \cdot \bar{\psi}^\mu &= \Lambda^\mu_\nu
    \bar{\psi}^\nu.
    \label{eq:poincareaction}
  \end{align}
\end{subequations}
On the antiholomorphic transverse fermions $\bar{\psi}^i$ the Lorentz
generators $J^{rs}$ are given by the zero modes of the currents
\begin{align}
  J^{rs}(z) &= \mathrm{i} :\bar{\psi}^r(z)\bar{\psi}^s(z):.
  \label{eq:lorentzcurrents}
\end{align}
Comparing equations (\ref{eq:ymcurrents}) and (\ref{eq:lorentzcurrents})
it becomes apparent that the formal treatment of both subsectors
is very similar.

The invariance under local superdiffeomorphisms leads to first class
constraints that are implemented in the BRST quantized theory via the
cohomology of the BRST operator $Q$. Representatives of the cohomology
groups in light cone gauge can be found by solving the mass equations
for the matter part $|\Psi\rangle_M$,
\begin{subequations}
  \begin{align}
    \left( L_0 - 1 \right) |\Psi\rangle_M &= 0 \\
    \left( \bar{L}_0 - \frac{1}{2} \right) |\Psi\rangle_M &= 0,
  \end{align}
  \label{eq:hstmass}
\end{subequations}
where $L_0$ and $\bar{L}_0$ are the zero modes of the holomorphic,
resp.~antiholomorphic Virasoro algebra. The ghost part of these states
are only important for string dynamics and are not considered here as
I am only concerned with the kinematics.

\subsection{Orbifold CFTs}
\label{sec:orbifoldcft}
\label{sec:orbifolds-gen}
The orbifolded CFT is constructed by keeping only states in the model that
are invariant under the action of a finite group $G$ of symmetries of the
state space. In addition this requires one to keep only those
observables that are invariant under the action of $G$. This
construction is described in detail by Dijkgraaf et al. in
\cite{Dijkgraaf}, Ginsparg \cite{Ginsparg} and Dixon et al. in
\cite{StringOrbifolds1,StringOrbifolds2}. 

There are three important consequences for model building. The first
one concerns the remnant symmetries. As the observable algebra
$\mathcal{A}^G$ of the model is reduced to the fixed points under $G$,
it is possible that Poincar\'e symmetry and Yang--Mills symmetry are
reduced by orbifolding, i.e. some generators are projected out. The
second one concerns the appearance of additional, twisted
representations of $\mathcal{A}^G$ that do not arise as subspaces of
the original state space. From the field theoretical point of view
they correspond to instanton sectors, i.e. one allows the fields to
be periodic up to the action of $G$. Equivalently, the worldsheet
theory has a discrete gauge symmetry $G$. Since observables are invariant
under $G$ all physical correlators are single valued thus leading to a
local worldsheet theory. The third and most important consequence is
the orbifold projection. Here, the spectrum is truncated in such a
way that only states invariant under orbifold group are kept. In the
following I will describe the generic orbifold projection.

The inclusion of twisted boundary conditions leads to a decomposition
of the total state space into twisted sectors $V_g$ one for each
group element $g \in G$,
\begin{align}
  V &= \bigoplus_{g\in G} V_g.
\end{align}

Let $|\phi,g\rangle \in V_g$ denote a state in which the fields obey
boundary conditions twisted by $g$. Let $h \in G$ be another orbifold
group element. Then, in the state $U(h)|\phi,g\rangle$ the fields obey
boundary conditions given by $hgh^{-1}$.
Thus, the operator $U(h)$ changed boundary conditions twisted by $g$
to boundary conditions twisted by $hgh^{-1}$, i.e. $U(h)|\phi,g\rangle \in
V_{hgh^{-1}}$. Because $G$ is a finite group, invariant states lie in
the image of the projection operator $P_{inv}$,
\begin{align}
  P_{inv} &= \frac{1}{|G|} \sum_{g\in G} U(g).
\end{align}
When analyzing an orbifold model that consists of several subsectors,
it is convenient to decompose the individual state spaces into
subspaces transforming irreducibly first and then use group
theoretical methods to find invariant states by combining irreducible
representations. Denote by $\chi_\alpha^g$ the character of the
irreducible representation $r_\alpha$ of the centralizer
$\mathcal{C}_G(g)$. The projector $P_\alpha^g$ is given by
\cite{Dijkgraaf}, $\iota \in G$ denotes the identity element,
\begin{align}
  P_\alpha^g &= \frac{\chi_\alpha^g(\iota)}{|\mathcal{C}_G(g)|}
  \sum_{h \in \mathcal{C}_G(g)} \chi_\alpha^g(h^{-1}) U(h).
  \label{eq:irrproj}
\end{align}
The projectors (\ref{eq:irrproj}) define a decomposition of the
twisted state space $V_g$ into subspaces $V_\alpha$ transforming
homogeneously under $\mathcal{C}_G(g)$,
\begin{subequations}
  \begin{align}
    V_g &= \bigoplus_\alpha r_\alpha \otimes V_\alpha \\
    r_\alpha \otimes V_\alpha &= P_\alpha^g(V_g).
  \end{align}
  \label{eq:irrdec}
\end{subequations}
States in $r_\alpha\otimes V_\alpha$ are said to transform
\emph{covariantly} under $\mathcal{C}_G(g)$. 

\subsection{Toroidal orbifolds}
A $\sigma$-model with target space being a toroidal orbifold provides
a simple example of an orbifold CFT and the geometry of the target
space is important to understand the spacetime theory. Some general
information on $\mathbb{Z}_2$ or $\mathbb{Z}_3$ orbifolds can be found
in \cite{StringOrbifolds2,FischerOrb,RatzStringy}.
A geometric orbifold $\mathcal{O} = M/G$ can be defined as a quotient
space of a smooth manifold $M$ by the action of a finite group
$G$. Two points $x_1$ and $x_2$ are identified if
\begin{align}
  x_1 \sim x_2 &\leftrightarrow x_1 = g\cdot x_2,
\end{align}
for some group element $g \in G$. The group $G$ is called the
\emph{orbifold group}. The group action may not be free,
i.e. there may exist fixed points. Geometric orbifolds fail to be
manifolds at the fixed points. 
Toroidal orbifolds are special cases of geometric
orbifolds. They are obtained by taking the manifold $M$
to be a torus. 
An $n$--dimensional torus $T^n$ is defined as the quotient of
$n$--dimensional Euclidean space $\mathbb{R}^n$ by a lattice $\Lambda$,
i.e.
\begin{align}
  T^n = \mathbb{R}^n/\Lambda,&\;\;\;\Lambda = \mathbb{Z}e_1 \oplus
  \ldots \oplus \mathbb{Z}e_n.
\end{align}
The lattice $\Lambda$ consists of all integer linear combinations of
the \emph{generators} $e_i \in \mathbb{R}^n$. The number of linearly
independent generators is called the \emph{rank of the lattice
  $\Lambda$}. 
A map $\sigma: \Lambda \rightarrow \Lambda$, which maps the lattice to
itself as a subset of $\mathbb{R}^n$ is called a \emph{lattice
  automorphism}. In this case, $\sigma$ induces a well-defined map
from the torus $T^n$ to itself. The map $\sigma: T^n \rightarrow T^n$
is called a \emph{torus automorphism}. The set of all torus
automorphisms constitutes a group under composition. A subgroup $P$ of
torus automorphisms is called a \emph{point group}. Every point group
is finite. The quotient space
\begin{align}
  \mathcal{O} &= T^n/P
\end{align}
is called a \emph{toroidal orbifold}.
To distinguish between inequivalent classes of geometric strings, it
is necessary to introduce the notion of a \emph{space group}. Instead
of constructing the orbifold $\mathcal{O}$ by dividing out first the
action of the lattice on $\mathbb{R}^n$ and then the action of the
point group on the torus $T^n$, it is possible to construct the
orbifold directly by dividing out the action of the \emph{space group}
$S$ on $\mathbb{R}^n$,
\begin{align}
  \mathcal{O} &\cong T^n/P \cong \mathbb{R}^n/S.
\end{align}
If there are no rototranslations the \emph{space group} $S = P \ltimes
\Lambda$ is a semidirect product of groups \cite{FischerOrb}.
The group $S$ acts on Euclidean space by translations and rotations as
($x \in \mathbb{R}^n$)
\begin{align}
  (g,\lambda) \cdot x &= g \cdot x + 2 \pi \lambda.
\end{align}

\section{Specification of the model}
\label{sec:model}
\subsection{Definition of the model}
From the geometric point of view, the orbifold model is completely
specified by the space group $S$. The space group $S$ is a
subgroup of the Poincar\'e group which is realized by
the generators $P^\mu$ and $J^{\mu\nu}$. A lattice element $\lambda \in \Lambda$
acts on Euclidean space by $(\iota,\lambda)
\cdot x = x + 2\pi\lambda$ and corresponds to a
translation. The quantum mechanical realization is provided by the
unitary operator
\begin{align}
  U(\iota,\lambda) &= \exp\left( 2\pi\imath \lambda_i P^i \right).
\end{align}
I assume for simplicity that there are no rototranslations, i.e. a
point group element $g \in P$ embeds into the space group $S$ as
$(g,0)$. In the heterotic string only the connected part of the Poincar\'e
group has an equivalent as a quantum mechanical operator. Therefore,
the rotation matrices $g$ must be orientation preserving, i.e.~$g \in
\SO(8)$. Every rotation matrix $g$ of finite order can be written in the
form
\begin{align}
  g &= \exp\left( 2\pi\imath \xi^i_g D_i^g \right),
  \label{eq:pointrep}
\end{align}
where the matrices $D_i^g$ generate rotations in mutually orthogonal
planes, hence, constitute a basis of the Lie algebra
$\mathfrak{so}(8)$. The four component vector $\xi_g$ is called the
\emph{twist vector} for the element $g$. Denote by $J_i^g$ the quantum
mechanical operators corresponding to $D_i^g$. Then, the geometric
action of $g$ is realized on the state space by the unitary operator
\begin{align}
  U(g,0) &= \exp\left( 2\pi\imath \xi^i_g J_i^g \right).
  \label{eq:pointgroupfermion}
\end{align}
There is a restriction on the choice of the twist vector $\xi_g$. The
quantum mechanical operators $U(g,0)$ constitute a representation of
the point group. Thus, the order of $U(g,0)$ and $g$ must be
identical. The spectrum of the heterotic string contains spinors and
the order of $U(g,0)$ can be doubled. In order to obtain a
well--defined action on spinors and $N\geq 1$ SUSY of spacetime, it is
sufficient to require \cite{StringOrbifolds2}
\begin{align}
  \sum_i \xi^i_g &\equiv 0 \mbox{ mod } 2.
\end{align}
The twist vector is not uniquely defined by the rotation matrix
$g$, because one can always add integers to each entry $\xi_g^i$
without changing $g$. Because the orbifold CFT can be different for
difference choices of the twist vector, it is part of the defining
data of the model. This phenomenon is also known from Abelian orbifold
model building and gives rise to models related by discrete torsion
\cite{MirageTorsion}. Although the choice of Cartan basis depends on
the sector $g$ the precise relation between two Cartan basis for
different elements is not important, because the basis is only used to
construct the appropriate twisted sector and the spectrum does not
depend on the other sectors.
To avoid a clash in notation, define a new (fermionic) twist vector
$\xi_g^{\mathrm{(F)}} = \xi_g$.

In addition to the geometric degrees of freedom the heterotic orbifold
has Yang--Mills degrees of freedom. These degrees of freedom are
realized by $32$ real holomorphic Majorana fermions $\Xi^I(z)$.
By equation (\ref{eq:ymcurrents}), there is a
representation of the group $\SO(16)\times \SO(16)$ on the fermionic
state space that is infinitesimally generated by operators
$\Omega^{IJ}$. An action of the point group is specified by giving a
homomorphism $\phi: \SO(8) \rightarrow \SO(16) \times \SO(16)$. Giving
the homomorphism 
$\phi$ is equivalent to selecting an $\mathfrak{so}(8)$-subalgebra
$\mathfrak{g}$ of $\mathfrak{so}(16)\oplus\mathfrak{so}(16)$.
To the generators $J_i^g$ there corresponds an element $K_i^g \in
\mathfrak{g}$ by the homomorphism $\phi$. The unitary maps on the state
space are then given by
\begin{align}
  U(g,0) &= \exp\left( 2\pi\imath (\xi_g^{\mathrm{(YM)}})^i K_i^g \right).
  \label{eq:pointgroupym}
\end{align}

\subsection{Twisted closed string boundary conditions}
\label{sec:orbifolds-strings-tbc}
Throughout this section the space group element
$(g,\lambda)$ is fixed. Denote by $n = \mbox{ord}(g)$ the order of the
rotation part.

\subsubsection{Free bosons with values in toroidal orbifolds}
The equations of motion, $\partial\bar{\partial} X^\mu(z,\bar{z}) = 0$,
have the same form as in the untwisted sector. But the boundary
conditions now read as
\begin{align}
  X( z e^{2\pi\imath}, \bar{z} e^{-2\pi\imath} ) &= g X(z,\bar{z}) +
  2\pi \lambda,
  \label{eq:orbibcfb}
\end{align}
where index contractions are implicit. The general solution to the
equations of motion subject to the orbifold boundary conditions
(\ref{eq:orbibcfb}) read
\begin{align}
  X(z,\bar{z}) &= x_0 - \imath a_0 \log{z} - \imath \bar{a}_0
  \log{\bar{z}} + \imath\sum_{k \in \mathbb{Z}\setminus\lbrace
    0\rbrace} \left( a_{\frac{k}{n}} 
  \frac{z^{-\frac{k}{n}}}{\frac{k}{n}} +
  \bar{a}_{\frac{k}{n}}\frac{\bar{z}^{-\frac{k}{n}}}{\frac{k}{n}} \right).
\end{align} 
The orbifold boundary conditions lead to restrictions
on the parameters $a_k$, $\bar{a}_k$ and $x_0$,
\begin{subequations}
\begin{align}
  \label{eq:5141}
  g x_0 + 2\pi \lambda &= x_0 + 2\pi (a_0 - \bar{a}_0), \\
  g a_{\frac{k}{n}} &= e^{-\frac{2\pi\imath k}{n}} a_{\frac{k}{n}}\mbox{ and } \\
  g \bar{a}_{\frac{k}{n}} &= e^{\frac{2\pi\imath k}{n}} \bar{a}_{\frac{k}{n}}.
\end{align}
\label{eq:fbconstraints}
\end{subequations}
The constructing elements $(\iota,\lambda)$ correspond to winding
modes and are already present on the torus. This suggests to introduce
the quantities
\begin{subequations}
  \begin{align}
    w &:= \frac{1}{2} ( a_0 - \bar{a}_0 )\mbox{ and } \\
    p &:= \frac{1}{2} ( a_0 + \bar{a}_0 ).
  \end{align}
  \label{eq:pwdef}
\end{subequations}
$w$ is called the \emph{winding vector} of the string and $p$ the
\emph{momentum} of the string. The latter name is motivated by the fact
that in the quantum theory the corresponding operator generates
translations in spacetime. Equations (\ref{eq:fbconstraints})
require invariance of both the winding vector $w$ and
the momentum $p$ under the action of the point group element $g$, just
set $k = 0$ there and use the definitions (\ref{eq:pwdef}).

Equation (\ref{eq:pointrep}) represents the point group element $g$ as
rotations in mutually orthogonal planes defined by the matrices
$D_i^g$. An orthogonal change of basis leaves the bosonic operator algebra
invariant, so that without loss of generality it is assumed that $g$
is blockdiagonal,
\begin{align}
  g &= \mbox{diag}\left(
  \mathcal{D}(2\pi\xi_g^1),\ldots,\mathcal{D}(2\pi\xi_g^4) \right),
  \label{eq:rotnormalform}
\end{align}
where $\mathcal{D}(\alpha)$ is a $2\times 2$ rotation matrix about
$\alpha$ clockwise.
The representation (\ref{eq:rotnormalform}) allows to analyze the
constraints (\ref{eq:fbconstraints}) separately for each plane
$D_g^i$. If $\xi_g^i \in \mathbb{Z}$, then the modes $a_{\frac{k}{n}}$
and $\bar{a}_{\frac{k}{n}}$ are integer moded and there the momentum
$p$ and center of mass coordinate $x_0$ are unconstrained. The winding
$w$ must be equal to the translation $\lambda$, i.e. $2 w =
\lambda$. The solutions are the same as for the torus. These planes
are called \emph{fixed tori}.

If the point group acts non--trivially on the plane, i.e. $\xi_g^i
\not\in\mathbb{Z}$, the constraints (\ref{eq:fbconstraints}) become
non-trivial. Denote by $X^1$ and $X^2$ the coordinate directions of
the plane $D_g^i$ and introduce complex coordinates by virtue of
\begin{subequations}
\begin{align}
  Z(z,\bar{z}) &= X^1(z,\bar{z}) + \imath X^2(z,\bar{z}) \\ 
  Z^*(z,\bar{z}) &= X^1(z,\bar{z}) - \imath X^2(z,\bar{z}).
\end{align}
\end{subequations}
The corresponding mode operators are denoted by $A_{\frac{k}{n}}$ and
$\bar{A}_{\frac{k}{n}}$, the center of mass coordinate by $Z_0$. In complex
coordinates, the action of $g$ is diagonal and reads as
\begin{align}
  g Z &= e^{-2\pi\imath\xi^i_g} Z.
\end{align}
The constraint equations (\ref{eq:fbconstraints}) can be solved and
read as
\begin{subequations}
  \begin{align}
    Z_0 &= \frac{2\pi}{1 - e^{-2\pi\imath\xi_g^i}} \lambda \\
    \xi_g^i - \frac{k}{n} \not\in\mathbb{Z} &\rightarrow A_{\frac{k}{n}} = 0 \\
    \xi_g^i + \frac{k}{n} \not\in\mathbb{Z} &\rightarrow
    \bar{A}_{\frac{k}{n}} = 0.
  \end{align}
  \label{eq:spacegroupfpequations}
\end{subequations}
There is no momentum $p=0$ and winding $w=0$ and the center of mass
coordinate is fixed to a point, the string is localized at
$Z_0$. The points $Z_0$ are called \emph{fixed points}. Twisted
strings wind themselves around the singularities\footnote{In geometric
orbifolds there are always singularities at the fixed points. This
follows from the fact that parallel transport of tangent vectors
around them has non trivial holonomy, but the geometry is flat near
the fixed points.} of the
metric. Moreover, the oscillators in the direction of this plane are
fractionally moded. Notably, there is a shift $\delta c$
in the zero--point energy of this twisted sector or equivalently, the
ground state in the twisted sector has non--zero conformal
weight. This shift reads as \cite{StringOrbifolds2}
\begin{align}
  \delta c &= \frac{1}{2} \sum_{i=1}^4 \eta_i \left( 1 - \eta_i
  \right),
  \label{eq:bosonshift}
\end{align}
where $\eta_i \in \xi_g^i + \mathbb{Z}$ with $0\leq \eta_i < 1$.

\subsubsection{Free fermions with values in orbifolds}
The worldsheet SUSY current is kept invariant under the action of the
orbifold group if the action on the antiholomorphic fermions is
defined via equation (\ref{eq:poincareaction}). This leads to twisted
sectors, where the fields $\psi(\bar{z})$ obey
\begin{align}
  \psi(\bar{z} e^{-2\pi\imath}) &= g \psi(\bar{z}).
\end{align}
In the fermionic realization of Yang--Mills symmetry the point group
acts on the fermions $\Xi^I(z)$ by (\ref{eq:pointgroupym}) and defines
an additional fermionic orbifold sector. The antiholomorphic fermion
sector and the Yang--Mills sector are structurally identical and only
the Yang--Mills sector is described in the following. For simplicity
it is assumed that the point group acts only on the first $16$
fermions. The orbifold is specified by giving a finite subgroup of
$\SO(16)$ transformations $g$ in the form (\ref{eq:pointgroupym}),
\begin{align}
  g &= \exp\left( 2\pi\imath (\xi_g^{\mathrm{(YM)}})^i K_i^g \right),
\end{align}
where $\xi_g^{\mathrm{(YM)}}$ is the twist vector for $g$ and $K_i^g$ form a basis for
a Cartan subalgebra of $\mathfrak{so}(16)$ and are normalized, such that
\begin{align}
  \frac{1}{2} \mbox{Tr} \left( K_i^g K_j^g \right) &= \delta_{ij}.
\end{align}
Consider the sector with boundary conditions twisted by $g$. The
fermionic mode algebra is left invariant by an
orthogonal transformation. Thus, it is no loss of generality to
assume that the matrix $g$ is block diagonal,
\begin{align}
  g &= \mbox{diag}\left(
  \mathcal{D}(2\pi(\xi_g^{\mathrm{(YM)}})^1),\ldots,\mathcal{D}(2\pi(\xi_g^{\mathrm{(YM)}})^8)
\right).
\end{align}
The fermions $\Xi^I(z)$ can be grouped into complex fermions
$\Psi^i(z)$ subject to the boundary conditions, cf. section
\ref{sec:enhanced}, 
\begin{align}
  \Psi^i(z e^{2\pi\imath} ) &= \pm e^{2\pi\imath(\xi_g^{\mathrm{(YM)}})^i}\Psi^i(z),
  \label{eq:orbi1}
\end{align}
where $\pm$ distinguishes between the NS or R sector.
The real fermions $\Xi^I(z)$ transform in the vector representation of
$\SO(16)$, so that $\Omega^{IJ}(z)$ transforms in the adjoint
representation. Thus, under the orthogonal transformation $g$ the
currents $\Omega^{IJ}(z)$ behave like
\begin{align}
  \Omega^{IJ}(z) &\overset{g}\rightarrow  g^I_R g^I_S \Omega^{RS}(z).
\end{align}
As for the free boson, twisted boundary conditions imply the
existence of fractionally numbered modes. Rewrite the currents
$\Omega^{IJ}(z)$ is a Cartan--Weyl basis, $H^i(z)$ and $E_\alpha(z)$,
w.r.t.\  the Cartan subalgebra defined by $K_i^g$. Then, the twisted
boundary conditions imply that
\begin{subequations}
  \begin{align}
    H^i(ze^{2\pi\imath}) &= H^i(z) \\
    E_\alpha(ze^{2\pi\imath}) &= e^{2\pi\imath \xi_g^{\mathrm{(YM)}} \cdot \alpha} E_\alpha(z).
  \end{align}
\end{subequations}
Only currents with $\xi_g^{\mathrm{(YM)}} \cdot \alpha \in \mathbb{Z}$ for all
roots $\alpha$ are periodic and, therefore, have zero--modes. Denote
the Lie algebra generated by the zero--modes by $\mathfrak{g}$. The
twisted state space carries only representations of $\mathfrak{g}$
instead of $\mathfrak{so}(16)$. 

\section{Projecting onto invariant states}
\label{sec:projecting}
\label{sec:orbiprojection}
\label{sec:orbifolds-strings-projection}
Constructing an orbifold model requires the removal of states not
invariant under the orbifold group. Geometric orbifold models are
defined by a space group. The methods presented in
section \ref{sec:orbifolds-gen} require that the acting group is
finite. However, the space group is infinite. By the
space group composition law every space group
element $(g,\lambda)$ can be decomposed as follows:
\begin{align}
  (g,\lambda) &= (g,0) \circ (\iota,g^{-1}\lambda).
  \label{eq:sgdec}
\end{align}
It is possible to arrive at a state space that is invariant under
the whole space group, by projecting onto states that are
invariant under the translations subgroup. The translation subgroup
does not act on the bosonic oscillators. Therefore, it is sufficient
to require its action on highest weight states with momentum $p$ and
winding $w$ to be trivial, i.e.
\begin{align}
  U(\iota,\lambda) | p, w \rangle &= e^{2\pi\imath p\cdot\lambda}|p,w\rangle,
\end{align}
where $\lambda \in \Lambda$ is an arbitrary lattice vector. The state
is invariant under translations if the phase factor is equal to $1$
for every lattice translation. Equivalently, the momentum $p$ must lie
in the dual lattice $\Lambda^*$. This \emph{quantization of momentum} is the
only restriction that arises from invariance under the
translations.
The problem of constructing a CFT invariant under an infinite group
has been reduced to constructing an orbifold CFT. For constructing
invariant states in the sector twisted by $g \in P$ it is enough to
project onto states that are invariant under the action of the
centralizer $\mathcal{C}_P(g)$. A general state $|\Psi\rangle$ in
the full state space is a tensor product of states from every
subsector, i.e.
\begin{align}
  |\Psi\rangle &= |\psi,b\rangle \otimes |\psi,f\rangle \otimes
  |\psi,\mathrm{YM}\rangle,
  \label{eq:genstate}
\end{align}
where the state $|\psi,b\rangle$ is from the bosonic sector, the state
$|\psi,f\rangle$ from the antiholomorphic fermion sector and
$|\psi,YM\rangle$ from the Yang--Mills sector.

If the centralizer is Abelian, then it is always possible to find a
basis in the state space, on which the centralizer elements $U(g) =
U(g,0)$ act diagonal. The reason is their representation theory:
abelian groups only have one-dimensional irreducible
representations, i.e. the group acts via multiplication by phase
factors there. However, 
if the centralizer is non--Abelian, then there are multidimensional
irreducible representations. The state space $V_i$ of each subsector
can be decomposed into parts transforming irreducibly, i.e.
\begin{align}
  V_i &= \bigoplus_{\alpha} r_\alpha \otimes V_{i,\alpha},
  \label{eq:decompose}
\end{align}
where $r_\alpha$ denotes an irreducible representation of the
centralizer and $V_{i,\alpha}$ is a subspace that is invariant under the
centralizer. Invariant states (\ref{eq:genstate}) are then
constructed, by taking states from $V_{i,\alpha}$ for each subsector and
combine them to an invariant state. 

\subsection{Remnant symmetries}
The orbifold observable algebra $\mathcal{A}^G$ contains only
observables of the original theory that are invariant under the whole
orbifold group. States represent this observable algebra. In the
original heterotic string, there are two important symmetries:
\begin{enumerate}
\item \emph{Poincar\'e symmetry} generated by the
  momentum operators $P^\mu$   and the rotation generators
  $J^{\mu\nu}$. The space group acts   non--trivially on the rotation
  generators and it is possible   to project out some of them out by
  restricting to the algebra of   invariant observables. Lorentz
  symmetry is reduced by orbifolding to   a subgroup $L \subset \SO(8)$,
  \begin{align}
    \SO(8) &\xrightarrow{\mathrm{orbifolding}} L \subset \SO(8).
  \end{align}
  
  For phenomenological reasons the point group
  acts only on six space--like coordinate directions. Notably, the
  point group leaves the four dimensional \emph{helicity operator} $J_{12}$
  invariant. Massless particles in four dimensions are classified
  according to irreducible representations of the stabilizer group of
  the standard momentum $(k,k,0,0)$. The helicity operator $J_{12}$
  generates rotations in transverse four--dimensional Minkowski
  space. Its eigenvalues determine the helicity of the particle.
  While $J_{12}$ is always fixed by the point group, there can be additional unbroken
  generators $J_{ij}$ from the compact directions. At first, $J_{ij}$
  seems to generate a continuous symmetry. But the models have to be
  invariant under lattice translations. Thus, only rotations generated
  by $J_{ij}$ that leave invariant the torus lattice, i.e. are torus automorphisms, are
  symmetries of the model. Because the torus automorphism group is
  always finite, the unbroken symmetry is always a discrete group. These
  symmetries need not to commute with the four--dimensional supercharge and
  are then called \emph{$R$ symmetries} \cite{Lee}.
  Let $R$ be a rotation that commutes with all point group
  elements and leaves the lattice invariant. $U(R,0)$ can map
  between different space group conjugacy classes and, hence, changes the
  localization of the string. Consider a string with constructing
  element $(g,\lambda)$, then
  \begin{align}
    U(R,0) U(g,\lambda) U(R,0)^{-1} &= U(g,R \lambda).
  \end{align}

  Because $R$ is a lattice automorphism, $R\lambda$ is again a lattice
  vector. But $(g,R\lambda)$ and $(g,\lambda)$ need not to be
  conjugated in $S$. Compatibility with the orbifold geometry
  requires in addition that $U(R,0)$ does not change the conjugacy
  classes.

\item \emph{Yang--Mills symmetry} is generated by the zero-modes of
  the currents $\Omega^{rs}(z)$ and the spin fields $S^\alpha(z)$. The
  orbifold projection projects some of them out. The invariant
  zero--modes generate a subalgebra $\mathfrak{g}_0 \subset
  \mathfrak{e}_8\otimes\mathfrak{e}_8$. The Yang--Mills gauge group of
  the effective, four--dimensional theory is broken to a subgroup $G_0$,
  \begin{align}
    \E_8\times \E_8 &\rightarrow G_0.
  \end{align}

  Physical states in the orbifold model transform
  in representations of $\mathfrak{g}_0$. These representations
  correspond to their charges under the remaining Yang--Mills
  symmetry. 
\end{enumerate}

\subsection{Analysis of the boson sector}
Sectors with non--trivial winding number $w$ are generically
massive. Because I am only interested in the effective massless
theory, I only consider the generic case henceforth. By equation
(\ref{eq:fbconstraints}) the
momentum $p$ of the string has to vanish in directions that do not
belong to fixed tori. Along fixed tori, the string is free to move, but
the momentum can only have discrete values. If the point group element
$g$ has a fixed point on the orbifold, the string is localized there.

Naively, there is one conjugacy class for each fixed point on the
torus. Naive conjugacy classes correspond to orbits
$[(g,\lambda)]_\Lambda$ of the space group under conjugation by the
translation subgroup, i.e.
\begin{align}
  [(g,\lambda)]_\Lambda &= \lbrace (\iota,\mu)\circ(g,\lambda) \circ
  (\iota,-\mu), \mu \in \Lambda \rbrace. 
\end{align}
Let's call those orbits $\Lambda$--orbits. The naive picture is true
if the representative $(g,\lambda)$ is mapped into its own
$\Lambda$--orbit under conjugation by an element of the point group.
But, it can also
happen that $(g,\lambda)$ is mapped into a different
$\Lambda$--orbit. In this case, the two orbits are identical under
the whole space group and the two fixed points have to be
identified. Identification of fixed points are
realized, e.g. in a $\mathbb{Z}_4$ orbifold \cite{Ibanez1}.
To each point group element there can correspond several space group
conjugacy classes. Each of them describes strings localized at
different fixed points. From the dynamical point of view they are
identical. However, by introducing Wilson lines, it is possible to
change the spectrum independently.
For each conjugacy class $f$ there is a highest
weight state $|0,0;f\rangle$, where the momentum and the winding have
been set to zero. The highest weight state is invariant under
rotations and, therefore, also invariant under the centralizer.

Consider the action of an element $h \in P$, commuting with $g$, on
the oscillators $a^i_{\frac{k}{n}}$. The element $h$ acts on them by
conjugation,
\begin{align}
  U(h) a^i_{\frac{k}{n}} U(h)^{-1} &= h^i_j a^j_{\frac{k}{n}}.
\end{align}
Because $h$ and $g$ commute, coordinate directions in which $g$ acts
differently are not mixed. Notably, 
fixed tori are mapped to fixed tori by $h$. Lorentz symmetry in the
transverse spacetime directions is reduced by 
the compactification to a subgroup $L \subset \SO(8)$. Similar to the
general case described in (\ref{eq:decompose}), the oscillators $a_{\frac{k}{n}}^i$ can be
decomposed into irreducible representations of both the residual
Lorentz group $L$ and the centralizer $\mathcal{C}_P(g)$.

The conformal weight of the fractionally moded oscillators is found by
the fact that $\mathrm{i}\partial X^\mu(z,\bar{z})$ is a primary
field. The commutator with $L_0$ evaluate to
\begin{align}
  [L_0, a_{\frac{k}{n}}^i ] &= -\frac{k}{n} a_{\frac{k}{n}}^i.
\end{align}
Applying $a_{-\frac{k}{n}}^i$ to a state raises its conformal weight
by $\frac{k}{n}$.
In the right--moving sector similar considerations can be made. This
time the oscillators $\bar{a}^i_{\frac{k}{n}}$ transform in the
complex conjugate representation, by equation (\ref{eq:fbconstraints}).

\subsection{Analysis of the Yang--Mills sector}
\label{sec:ymanalysis}
\label{sec:orbianalysis}
Consider a sector with boundary conditions twisted by
$g$. In the bosonized description the
fermions $\Psi^i(z)$ are realized by bosons $\phi^i(z)$. The momentum
of the bosons corresponds to the eigenvalues of the operators
$H^i_0$ and is constrained to values in a shifted $\E_8$ root lattice,
\begin{align}
  (p^i) &\in \xi_g + \Lambda_{\E_8}.
  \label{eq:shiftedmomentum}
\end{align}

States have to be invariant under the action of the centralizer
$\mathcal{C}_P(g)$. In the definition (\ref{eq:pointgroupym}) of the
action of the point group there is an ambiguity. In general, the
centralizer is only represented projectively \cite{Dijkgraaf}.

\subsection{Finding representations of $\mathfrak{g}$ at a given
  conformal weight.}
The unprojected state space is independent of the orbifold projection and
of the definition of the operators $U(h,0)$. I use the bosonized
description to find the states and representations of
$\mathfrak{g}_0$ at a given conformal weight. Recall that
$\mathfrak{g}$ is the Lie algebra generated by the zero--modes of the
currents $\Omega^{rs}(z)$.

An algorithm to find all irreducible representations
of the algebra $\mathfrak{g}$ at a conformal weight $L_0 = A$ is given
now. A state is characterized by its momentum $(p^i)$ and the oscillator numbers
$N_{osc}^i, i=1,\ldots,8$. By equations (\ref{eq:akmmass}) and (\ref{eq:shiftedmomentum}) one
has to find $\E_8$ lattice vectors $\alpha$ and non--negative integers
$n^i$, such that
\begin{align}
  A &= \frac{1}{2} (\alpha + \xi_g^{\mathrm{(YM)}})^2 + \sum_{i=1}^8 n_i.
  \label{eq:leq}
\end{align}

The Diophantine problem (\ref{eq:leq}) is equivalent to finding
lattice points inside a sphere of radius $\sqrt{2A}$ in $\mathbb{R}^8$
shifted by the vector $\xi$. For small $A$ it can be solved
numerically, e.g. by using GAP \cite{GAP}. Denote by $W_A$ the
collection of momenta $(p^i) = \alpha + \xi_g^{\mathrm{(YM)}}$ that solve
(\ref{eq:leq}) including degeneracies. Since the state space
represents the zero--mode algebra $\mathfrak{g}$ and the momenta are
the eigenvalues of the operators $H_0^i \in \mathfrak{g}$, the
collection $W_A$ must split into collections of weights of
irreducible representations of $\mathfrak{g}$. Equivalently, the
eigenspace $V_A \subset V$ of $L_0$ for eigenvalue $L_0 = A$
decomposes into irreducible representations of $\mathfrak{g}$ as
\begin{align}
  V_A &= \bigoplus_i L(\lambda_i,q_i),
  \label{eq:statedec}
\end{align}

where $\lambda_i \in W_A$ are highest weight vectors of
$\mathfrak{g}_s$ and $q_i$ is a tuple of charges
$Q_i$. $\mathfrak{g}_s$ is the semisimple part of $\mathfrak{g}$. The
charges $Q_i$ form a basis of the $\mathfrak{u}(1)$ parts and are
linear combinations of the generators $H_0^i$.

\subsection{Projecting onto $\mathcal{C}_P(g)$ covariant states.} 
I this section I decompose twisted state space into irreducible
representations of the centralizer and the invariant algebra
$\mathfrak{g}_0$.
Let be $h \in \mathcal{C}_P(g)$. If $X \in \mathfrak{g}$ is arbitrary,
it is left invariant by $g$, i.e. $U(g) X U(g)^{-1} = X$, by definition of
$\mathfrak{g}$. But the element $U(h) X U(h)^{-1}$ is also left invariant by
$g$, because $g$ and $h$ commute. Hence, the operator $U(h)$ induces a Lie
algebra automorphism $\mathfrak{g} \rightarrow
\mathfrak{g}$. Notably, it maps irreducible representation of
$\mathfrak{g}$ into themselves and it suffices to restrict to a
representation $L(\lambda_i,q_i)$ from the decomposition
(\ref{eq:statedec}).

Consider a Cartan--Weyl basis $I^i = H^i$, $F_\alpha$ of
$\mathfrak{g}$ and fix a choice of positive roots. If there are
$\mathfrak{u}(1)$ factors, appended them to the basis elements
$I^i$ of the Cartan subalgebra. States of the form
$F_{-\alpha_1}\ldots F_{-\alpha_n}|\lambda_i,q_i\rangle$ ($\alpha_i$
positive roots) span the subspace $L(\lambda_i,q_i)$ for a highest
weight vector $|\lambda_i,q_i\rangle$.

The operator $U(g)U(h)U(gh)^{-1}$ acts as the identity on
$\mathfrak{g}$. Therefore, its action on the highest weight
representation $L(\lambda_i,q_i)$ is completely determined by
its action on the highest weight state
$|\lambda_i,q_i\rangle$. However, in a highest weight module the
weight space for $\lambda_i$ is always
one--dimensional \cite{Fulton}. Thus, using Schur's lemma and
unitarity of $U(g)$ the action has to be multiplication by a phase
factor,
\begin{align}
  U(g)U(h)U(gh)^{-1} = e^{\imath\alpha(g,h)}.
\end{align}

If the phase factor $e^{\imath\alpha(g,h)}$ is non--trivial, the
centralizer is projectively represented. By the remark below it can be
redefined to a linear representation by multiplying with an appropriate
phase function $f(g)$,
\begin{align}
  U(g) &\rightarrow f(g)^{-1} U(g).
\end{align}

The projection onto irreducible representations of $\mathcal{C}_P(g)$
is accomplished by using the projection formulas
(\ref{eq:irrproj}). According to (\ref{eq:irrdec}), the representation
$L(\lambda_i,q_i)$ decomposes as
\begin{align}
  L(\lambda_i,q_i) &= \bigoplus_{\alpha} r_\alpha \otimes V_\alpha,
\end{align}

where $r_\alpha$ are irreducible representations of the
centralizer. Because $L(\lambda_i,q_i)$ are irreducible, the subspaces
$V_\alpha$ are not representations of $\mathfrak{g}$ anymore, in
general. In the orbifold model only the action of the modes 
invariant under the whole orbifold group $P$ is well--defined in all
sectors. Hence, the subspaces $V_\alpha$ carry representations of
$\mathfrak{g}_0$ and can be decomposed into irreducible
representations of it.

\subsection{Analysis of the fermion sector}
Structurally, the antiholomorphic fermion sector is the same as the
Yang--Mills sector. The only formal difference is that the number of
complex fermions this time is $4$ instead of $8$ (or $16$ if both
$\E_8$ factors are involved) and there is no enhanced AKM symmetry. In
the bosonized description the momenta $(q^i)$ are given this time by
(\ref{eq:fermionmomenta}), i.e.
\begin{align}
  (q^i) &\in -\xi_g^{\mathrm{(F)}} + \Lambda_{\SO(8)}\mbox{ for the NS--sector and}\\
  (q^i) &\in -\xi_g^{\mathrm{(F)}} + \lambda_s + \Lambda_{\SO(8)}\mbox{ for the
    R--sector,}
  \label{eq:tmp1}
\end{align}
where $\lambda_s$ is a highest weight vector for the spinor
representation of $\SO(8)$ and $\xi_g^{\mathrm{(F)}}$ denotes the twist vector for the
$g$--twisted sector. The minus signs in front of the shift vectors in
(\ref{eq:tmp1}) take into account that the representations in the
antiholomorphic sector are complex conjugated. Finding representations
at a given conformal weight and the projection on
$\mathcal{C}_P(g)$ covariant states goes along the same lines as for
the Yang--Mills sector.

\subsection{Analysis of the full state space}
The state space of the full orbifold CFT is a subspace
of the tensor product of the bosonic, fermionic and the Yang--Mills
sector that is invariant under the whole space group $S$. Not all
states in the orbifold state space are physical. They have to satisfy
the \emph{mass equations} (\ref{eq:hstmass}),
\begin{subequations}
\begin{align}
  \left( L_0 - 1 \right) |\Psi\rangle_M &= 0 \\
  \left( \bar{L}_0 - \frac{1}{2} \right) |\Psi\rangle_M &= 0
\end{align}
\end{subequations}
on the matter parts of physical states. The matter part consists of
the bosonic, fermion and the Yang--Mills sector only. Since the total
matter stress energy tensor is the sum of the stress energy tensors of
the individual sectors, the mass equations in the $g$--twisted sector
can be rewritten as
\begin{subequations}
  \begin{align}
    \left( L_0^{\mathrm{(YM)}} + \frac{1}{2} (p+w)^2 + N_{\mathrm{osc}} + \delta c -
      \frac{1}{2} m^2 - 1 \right) |\Psi\rangle_M &= 0 \\
    \left( \bar{L}_0^{\mathrm{(fermion)}} + \frac{1}{2} (p-w)^2 + \bar{N}_{\mathrm{osc}} + \delta c -
      \frac{1}{2} m^2 - \frac{1}{2} \right) |\Psi\rangle_M &= 0,
  \end{align}
  \label{eq:hstmass2}
\end{subequations}

where $m^2$ is the mass in four dimensions, $p \in \Lambda^*$ the
internal momentum and $2w \in \Lambda$ is the winding
vector. $N_{osc}$ and $\bar{N}_{osc}$ are the contribution of the
bosonic oscillators from both compactified and uncompactified
directions in the holomorphic, resp. antiholomorphic sector
of the model. $\delta c$ is the shift of zero--point energy
(\ref{eq:bosonshift}) in the twisted sectors. Because of the symmetric
action of the orbifold group, it is the same in both chiral parts.
The operators $L_0^{\mathrm{(YM)}}$ and $\bar{L}_0^{\mathrm{(fermion)}}$ evaluate to the
conformal weight of the Yang--Mills part and the fermionic part,
respectively, and can be calculated by the methods presented above.

\section{Constraints from one-loop modular invariance}
In perturbative string theory a $g$ loop amplitude is obtained by
first calculating the CFT correlation function of appropriate vertex
operators on a genus $g$ Riemann surface and, then, by integrating the
result over the moduli space of punctured Riemann surfaces. In order
to yield well-defined amplitudes, the correlation functions have to be
single-valued on the moduli space. This condition is called modular
invariance. In this section I find the one-loop contribution to
the string free energy. This contribution is given by calculating the
zero-point function on a torus with modular parameter $\tau$. I then
calculate their behaviour under modular transformations and give
sufficient conditions for modular invariance. These conditions turn
out to be very similar to the one known from Abelian orbifold model
building and reduce to them upon choosing an Abelian point group.

The state space of the CFT decomposes into sectors labelled by the
conjugacy classes $[g]$ of the point group $P$. The partition function
then takes the form
\begin{align}
  \mathcal{Z}(\tau,\bar{\tau}) &= \sum_{[g]}
  \frac{1}{|\mathcal{C}_P(g)|} \sum_{h \in \mathcal{C}_P(g)}
  \mathrm{Tr}_{V_g} \left( U(h) q^{L_0 - c/24} \bar{q}^{\bar{L}_0 -
      \bar{c}/24} \right) = \notag \\
  &= \sum_{[g]}
  \frac{1}{|\mathcal{C}_P(g)|} \sum_{h \in \mathcal{C}_P(g)}
  \mathcal{Z}_{g,h}(\tau,\bar{\tau}),
\end{align}
where $q = e^{2\pi\mathrm{i}\tau}$ is the nome and $U(.)$ denotes the
(projective) representation of $h \in \mathcal{C}_P(g)$ on the
subspace $V_g$. The partition function can be rewritten as
\cite{Dijkgraaf},
\begin{align}
  \mathcal{Z}(\tau,\bar{\tau}) &= \frac{1}{|P|} \sum_{[g,h] = \iota}
  \mathcal{Z}_{g,h}(\tau,\bar{\tau}).
\end{align}
Note that compared to \cite{Dijkgraaf} our definition of the objects
$\mathcal{Z}_{g,h}(\tau,\bar{\tau})$ contains already possible
discrete torsion phases. They are defined as the partition function of
the CFT on the torus with boundary conditions $g$ in the $a$-direction
and $h$ in the $b$-direction, where $a$ and $b$ denote a canonical
basis of the first homology group of the torus. I call
$\mathcal{Z}_{g,h}(\tau,\bar{\tau})$ the (non-chiral) orbifold characters.

Since the total states space is a tensor product of four different
CFTs, the orbifold characters decompose into a product of four
characters,
\begin{align}
  \mathcal{Z}_{g,h}(\tau,\bar{\tau}) &=
  \mathcal{Z}^{\mathrm{B}}_{g,h}(\tau,\bar{\tau}) \mathcal{Z}^{\mathrm{YM}}_{g,h}(\tau)
  \mathcal{Z}^{\mathrm{fermion}}_{g,h}(\bar{\tau})
  \mathcal{Z}^{\mathrm{rest}}_{g,h}(\tau,\bar{\tau}).
\end{align}
The character $\mathcal{Z}^{\mathrm{rest}}_{g,h}(\tau)$ stems from the
ghosts and the uncompactified directions. It is invariant under
$S$-modular transformations and cancels some
$e^{-\frac{\mathrm{i}\pi}{12}}$ factors arising from $T$-modular
transformations. $\mathcal{Z}^{\mathrm{B}}_{g,h}(\tau,\bar{\tau})$
describes the contribution from the bosonic part and
$\mathcal{Z}^{\mathrm{fermion}}_{g,h}(\bar{\tau})$ and
$\mathcal{Z}^{\mathrm{YM}}_{g,h}(\tau)$ describe the contributions
from the fermionic and the Yang--Mills sector, respectively. The
latter characters are chiral, because the associated CFT is
chiral. It has been shown in \cite{Vafa} that
$\mathcal{Z}^{\mathrm{B}}(\tau,\bar{\tau})$ is modular invariant at
one-loop in arbitrary orbifold compactifications of the heterotic
string. Moreover, there it is also shown that the difference between
the holomorphic and antiholomorphic fermionic ground state energies
have to be equal modulo $1$. This is the same condition that follows
by directly calculating the fermionic contributions to the partition
function.

\subsection{The chiral orbifold characters}
The calculation of chiral orbifold characters for the fermionic sector
and the Yang--Mills sector is similar. Henceforth I only
describe the calculation for the Yang--Mills sector and just state the
results for the fermionic part.

Let $h \in P$ commute with $g$ and consider the action $U(h)$ in the
twisted sector $V_g$. $U(h)$ is possibly a projective representation,
so that its action on the state space is only defined upto some phase
factor. So, a priori, the chiral orbifold characters are only defined
up to a phase factor. However, the action on the fields $\Xi^I(z)$ is
still well-defined. In the sector $V_g$ the real fermions $\Xi^I$ are
grouped into complex fermions $\Psi^i$ subject to the boundary
conditions, c.f.~appendix \ref{app:fermionicakm},
\begin{align}
  \Psi^i(z e^{2\pi\mathrm{i}}) &= \pm e^{2 \pi \mathrm{i}
    (\xi_g^{\mathrm{(YM)}})^i} \Psi^i(z),
\end{align}
where $\pm$ distinguishes between the NS and R sector.
Thus, the twisted state space $V_g$ is a tensor product of $8$
complex, twisted fermions as described in appendix
\ref{app:fermionicakm}. The choice of basis for the Cartan subalgebra
enters in the calculation through the grouping of the real fermions
into complex fermions and through the twist vector
$\xi_g^{\mathrm{(YM)}}$. Denote by $H_0^i$ the zero modes of the
Cartan currents. Because $g$ and $h$ commute, $h$ induces an
automorphism of the invariant subalgebra $\mathfrak{g} =
\mathfrak{so}(16)^g \subset \mathfrak{so}(16)$ and also of any
representation of $\mathfrak{g}$. Denote the exponential group of
$\mathfrak{g}$ by $G$. Because $h$ has finite order, it is conjugated
to an element in the maximal torus of exponentiated Cartan
elements, $R \in G$,
\begin{align}
  R U(h) R^{-1} &= \exp\left( 2\pi\mathrm{i}\zeta^i (H^i_0 - \xi_g^{\mathrm{(YM)}}) \right) = G_{\zeta}. 
\end{align}
For later convenience I added the constant factor $e^{-2\pi\mathrm{i}
  \zeta\cdot\xi_g}$ to $G_\zeta$. This factor does not change the
adjoint action of $h$ on the fermionic modes.
The operators $R$ have conformal weight $0$ and, hence, commute with
$L_0$. Thus, the chiral orbifold character for the Yang--Mills sector
can be written as
\begin{align}
  \mathcal{Z}_{g,h}^{\mathrm{(YM)}}(\tau) &= \mathrm{Tr}_{V_g} \left(
    G_\zeta q^{L_0 - c/24}\right).
\end{align}
Since the fields $\Psi^i$ transform in the vector representation of
$\mathfrak{so}(16)$, the operator $G_\zeta$ induces a gradation of type $\zeta$
on the fermionic Fock space and the chiral orbifold characters can be
evaluated explicitly in terms of Jacobi theta functions, see appendix
\ref{app:fermionicakm} for more details. Using the general characters
$\mathcal{Z}_{\xi,\zeta}$ introduced in appendix \ref{app:fermionicakm} the
chiral orbifold characters evaluate to
\begin{align}
  \mathcal{Z}_{g,h}^{(YM)}(\tau) &= \mathcal{Z}_{\xi_g^{\mathrm{(YM)}},\zeta_h^{g}}(\tau).
\end{align}
The vector $\zeta_h^g$ corresponds to the chosen $G_\zeta$. Note that
the choice of $\zeta$ is not unique. If we replace $\zeta
\rightarrow \zeta + \alpha$, where $\alpha \in
\Lambda_{\mathfrak{so}(16)}^*$ is a weight of $\mathfrak{so}(16)$, it
still induces the same automorphism
of $\mathfrak{g}$ but might act differently on states $| p \rangle \in
V_g$. However, a simple calculation shows that
\begin{align}
  G_{\zeta+\alpha}|p\rangle &= G_\zeta|p\rangle.
\end{align}
This means that the operator $G_\zeta$ is uniquely defined as are the
chiral orbifold characters. The additional factor in $G_\zeta$ ensures
that the twisted Fock vacuum transforms trivially under
$G_\zeta$. Additionally, one finds that, $\alpha \in \Lambda^*_{\mathfrak{so}(16)}$,
\begin{subequations}
\begin{align}
  \mathcal{Z}_{\xi_g^{\mathrm{(YM)}},\zeta_h^g + \alpha}(\tau)
  &=\mathcal{Z}_{\xi_g^{\mathrm{(YM)}},\zeta_h^g}(\tau) \\
  \mathcal{Z}_{\xi_g^{\mathrm{(YM)}}+\alpha,\zeta_h^g}(\tau)
  &= e^{-2\pi\mathrm{i}
    \alpha \cdot\zeta_g^h}\mathcal{Z}_{\xi_g^{\mathrm{(YM)}},\zeta_h^g}(\tau).
\end{align}
\end{subequations}
Formula (\ref{eq:fermionakmchars}) describing the behaviour of the orbifold
characters $\mathcal{Z}_{\xi,\zeta}(\tau)$ under modular transformations implies that 
\begin{subequations}
  \begin{align}
    \mathcal{Z}_{\xi_g^{(YM)},\zeta^g_h}(\tau)
    &\overset{T}{\rightarrow} e^{-\frac{8\imath\pi}{12}}
    e^{\imath\pi(\xi_g^{(YM)})^2}
    \mathcal{Z}_{\xi_g^{(YM)},\xi_g^{(YM)}+\zeta_h^g}(\tau) \\
    \mathcal{Z}_{\xi_g^{(YM)},\zeta_h^g}(\tau)
    &\overset{S}{\rightarrow}
    e^{-2\imath\pi \xi_g^{(YM)} \cdot \zeta_h^g}
    \mathcal{Z}_{\zeta_h^g,-\xi_g^{(YM)}}(\tau).
  \end{align}
  \label{eq:ymchartrans}
\end{subequations}
Recall that the operator $R$ diagonalizing $U(h)$ is contained in the
group $G = \exp\mathfrak{g}$ and commutes by definition with the
matrix $U(g)$. So, we see that $\zeta^g_{gh} = \zeta^g_h +
\xi_g^{\mathrm{(YM)}}$ up to a weight vector. This means that the set
of orbifold characters is closed under $T$-modular transformations.

The operator $R$ has fermion number zero and is an exponential of a
quadratic expression in the fermionic mode operators. This means that
it induces a Bogoliubov transformation on the twisted fermionic Fock
space: $\Xi^I \rightarrow R^{-1} \Xi^I R$. The transformed vacuum
$R^{-1}|0\rangle$ is the twisted vacuum of a theory with boundary
conditions $R h R^{-1}$ instead of $h$. Consequently, one obtains
\begin{align}
  \mathcal{Z}_{\zeta_h^g, -\xi_g^{\mathrm{(YM)}}}(\tau) &=
  \mathrm{Tr}_{R h R^{-1}} \left( q^{L_0 - c/24} U(g^{-1}) \right) =
  \mathrm{Tr}_{h} \left( R q^{L_0 - c/24} U(g^{-1}) R^{-1}\right)
  \notag \\
  &=
  \mathrm{Tr}_{h} \left( q^{L_0 - c/24} U(g^{-1})\right)
  =  \mathcal{Z}_{h,g^{-1}}^{\mathrm{(YM)}}(\tau).
\end{align}
Thus, the orbifold characters also close under $S$-modular
transformations. 

\subsection{Invariance of the full partition function}
Combining the results of the previous subsection one deduces the
following transformation rules for the non-chiral characters:
\begin{subequations}
\begin{align}
  \mathcal{Z}_{g,h}(\tau,\bar{\tau}) &\overset{T}{\rightarrow}
  e^{\pi\mathrm{i} \left( (\xi_g^{\mathrm{(YM)}})^2 -
      (\xi_g^{\mathrm{(F)}})^2 \right)}
  \mathcal{Z}_{g,gh}(\tau,\bar{\tau}) \\
  \mathcal{Z}_{g,h}(\tau,\bar{\tau}) &\overset{S}{\rightarrow}
  e^{-2\pi\mathrm{i} \left( \xi_g^{\mathrm{(YM)}} \cdot \zeta_h^g -
      \xi_g^{\mathrm{(F)}} \cdot \bar{\zeta}_h^g \right)}
  \mathcal{Z}_{h,g^{-1}}(\tau,\bar{\tau}),
\end{align}
\end{subequations}
where $\zeta_h^g$ denotes the vector $\zeta^i$ occuring in the
definition of $G_\zeta$ in the previous subsection. Similarly,
$\bar{\zeta}_h^g$ corresponds to the value for the antiholomorphic
fermionic sector. It now follows, that $T$-modular invariance requires
that the ground state energies for the holomorphic and antiholomorphic
fermions are equal modulo $N = \mathrm{ord}(g)$:
\begin{align}
  N \left( (\xi_g^{\mathrm{(YM)}})^2 -
      (\xi_g^{\mathrm{(F)}})^2 \right) &\equiv 0 \mbox{ mod } 2
\end{align}
If this condition is also fulfilled for $N = 1$, the model is strongly
modular invariant. Note that models with a standard embedding are
always modular invariant.

\section{The {\boldmath $S_3$} orbifolded heterotic string}
\label{sec:example}
In this section I demonstrate the methods developed in the previous
sections by explicitly constructing a heterotic orbifold model with
point group $S_3$. This geometry occurs in the $2262$nd
$\mathbb{Q}$--class in the classification of \cite{FischerOrb} and is
known to allow for $N=1$ SUSY. I find
the remnant symmetries and the
transformation properties of the massless single string
states. Finally, I discuss its untwisted geometric moduli.

\subsection{Properties and representations of the group {\boldmath
    $S_3$}}
\label{app:s3}
The symmetric group $S_3 = \lbrace \iota, \tau, \sigma, \tau\sigma, \sigma^2,
  \tau\sigma^2 \rbrace$ has six elements and is presented by
\begin{align}
  S_3 &= \langle \tau, \sigma | \tau^2 = \sigma^3 = (\tau\sigma)^2 =
  \iota \rangle.
  \label{eq:s3presentation}
\end{align}

\begin{table}[t]
  \begin{minipage}{6cm}
    \begin{center}
  \begin{tabular}{c||ccc|}
    & $[\iota]$ & $[\tau]$ & $[\sigma]$ \\[.3em]
    \hline\hline
    $\chi_{\mathrm{inv}}$ & 1 & 1 & 1 \\
    $\chi_{\mathrm{alt}}$ & 1 & -1 & 1 \\
    $\chi_{\mathrm{def}}$ & 2 & 0 & -1 \\[.3em]
    \hline
  \end{tabular}
  \end{center}
  \end{minipage}\hfill
  \begin{minipage}{8cm}
    \begin{center}
  \begin{tabular}{c||ccc|}
    $\otimes$ & $r_{\mathrm{inv}}$ & $r_{\mathrm{alt}}$ & $r_{\mathrm{def}}$ \\[.3em]
    \hline\hline
    $r_{\mathrm{inv}}$ & $r_{\mathrm{inv}}$ & $r_{\mathrm{alt}}$ & $r_{\mathrm{def}}$ \\
    $r_{\mathrm{alt}}$ &  & $r_{\mathrm{inv}}$ & $r_{\mathrm{def}}$ \\
    $r_{\mathrm{def}}$ &  &  & $r_{\mathrm{inv}} \oplus r_{\mathrm{alt}}\oplus r_{\mathrm{def}}$ \\[.3em]
    \hline
  \end{tabular}
  \end{center}
  \end{minipage}

  \caption{Character table and composition rules for the group
    $S_3$. The irreducible representation with character $\chi_{\mathrm{inv}}$,
    $\chi_{\mathrm{alt}}$ or $\chi_{\mathrm{def}}$ are called the trivial $r_{\mathrm{inv}}$,
    alternating $r_{\mathrm{alt}}$ or defining $r_{\mathrm{def}}$ representation,
    respectively \cite{GAP,Fulton}.}
  \label{tab:s3tab}
\end{table}

In this notation $S_3$ is the free group in two generators,
$\tau$ and $\sigma$, with the only relations given on the right--hand
side.
$\tau$ has order $2$ and $\sigma$ has order $3$. By the identification
$\tau \equiv (1,2)$ and $\sigma \equiv (1,2,3)$ the familiar
realization of the symmetric group as permutations of three objects,
$1$, $2$ and $3$, is recovered. The presentation notation emphasizes the
relations among the group elements rather than some specific realization.
The group $S_3$ splits into three conjugacy classes,
  \begin{align}
  [\iota] = \lbrace \iota \rbrace,\;
  [\tau] = \lbrace \tau, \tau\sigma, \tau\sigma^2 \rbrace,\;&
  [\sigma] = \lbrace \sigma, \sigma^2 \rbrace.
\end{align}

The number of irreducible representations of a finite group is equal
to the number of conjugacy classes: one two dimensional irreducible
representation and two one dimensional
representations. The latter satisfy $r(\tau) = \pm 1$ and
$r(\sigma) = 1$. The first 
choice of sign denotes the \emph{invariant} or \emph{trivial
  representation} $r_{\mathrm{inv}}$, the second is called the
\emph{alternating representation} $r_{\mathrm{alt}}$. The third irreducible
representation $r_{\mathrm{def}}$ is called the \emph{defining
  representation}. It can be realized by $\GL(2,\mathbb{Z})$ matrices
as
\begin{align}
  r_{\mathrm{def}}(\tau) = \left( \begin{array}{cc} 1 & -1 \\ 0 & -1
    \end{array} \right),\;\;r_{\mathrm{def}}(\sigma) = \left( \begin{array}{cc}
      0 & -1 \\ 1 & -1 \end{array} \right).
\end{align}

The character table together with the composition rules of
these irreducible representations can be found in table
\ref{tab:s3tab}. The projectors onto irreducible components in
terms of elements of the group ring $\mathbb{C} S_3$ are given,
according to equation (\ref{eq:irrproj}):
\begin{align}
  P_{\mathrm{inv}} &= \frac{1}{6} \left( \iota + \tau \right) \left( \iota +
    \sigma + \sigma^2 \right),\;\;
  P_{\mathrm{alt}} &= \frac{1}{6} \left( \iota - \tau \right) \left( \iota +
    \sigma + \sigma^2 \right),\;\;
  P_{\mathrm{def}} &= \frac{1}{3} \left( 2 \iota -
    \sigma - \sigma^2 \right).
\end{align}
\begin{table}[t]
 \begin{minipage}{6.5cm}
    \begin{center}
  \begin{tabular}{c||cc|}
    & $[\iota]$ & $[\tau]$  \\[.3em]
    \hline\hline
    $\chi_{+}$ & 1 & 1 \\
    $\chi_{-}$ & 1 & -1 \\[.3em]
    \hline
  \end{tabular} \hspace{5mm}
  \begin{tabular}{c||cc|}
    $\otimes$ & $r_{+}$ & $r_{-}$ \\[.3em]
    \hline\hline
    $r_{+}$ & $r_{+}$ & $r_{-}$  \\
    $r_{-}$ &  & $r_{+}$ \\[.3em]
    \hline
  \end{tabular}
  \end{center}
  \end{minipage}\hfill
 \begin{minipage}{8cm}
    \begin{center}
      \begin{tabular}{c||ccc|}
        & $[\iota]$ & $[\sigma]$ & $[\sigma^2]$  \\[.3em]
        \hline\hline
        $\chi_{0}$ & 1 & 1 & 1 \\
        $\chi_{+}$ & 1 & $\zeta^2$ & $\zeta$ \\
        $\chi_{-}$ & 1 & $\zeta$ & $\zeta^2$ \\[.3em]
        \hline
      \end{tabular}\hspace{5mm}
      \begin{tabular}{c||ccc|}
        $\otimes$ & $r_0$ & $r_{+}$ & $r_{-}$ \\[.3em]
        \hline\hline
        $r_0$ & $r_0$ & $r_+$ & $r_-$ \\
        $r_{+}$ & & $r_{-}$ & $r_{0}$  \\
        $r_{-}$ & &  & $r_{+}$ \\[.3em]
        \hline
      \end{tabular}
    \end{center}
  \end{minipage}

  \caption{Character table and composition rules for the groups
    $\mathbb{Z}_2$ (left) and $\mathbb{Z}_3$ (right). The irreducible
    representation of $\mathbb{Z}_2$ with character $\chi_{+}$ or
    $\chi_{-}$ will be called the trivial $r_{+}$ or
    alternating $r_-$ representation, respectively. For
    $\mathbb{Z}_3$, the irreducible
    representation with character $\chi_{0}$, $\chi_+$ or
    $\chi_{-}$ will be called the trivial $r_{0}$,
    defining $r_+$ representation or complex conjugated defining $r_-$
    representation, respectively. The symbol $\zeta =
    \exp(2\pi\imath/3)$ is a third root of unity.}
  \label{tab:z2ctab}
  \label{tab:z3ctab}
\end{table}
The centralizer of $\tau$ in $S_3$ is $\mathcal{C}_{S_3}(\tau) =
\langle \tau \rangle \cong \mathbb{Z}_2$ and for $\sigma$ it is given
by $\mathcal{C}_{S_3}(\sigma) = \langle \sigma \rangle \cong
\mathbb{Z}_3$. Both groups are cyclic and, hence,
Abelian. Their character tables and fusion rules can be found in table
\ref{tab:z2ctab}.

\subsection{The geometric orbifold}
\label{sec:s3orbifoldgeometry}
A toroidal orbifold is completely specified by its lattice $\Lambda$
of translations and the point group. $\Lambda$ is an
$\SU(3)\times\SU(3)\times\SU(3)$ lattice. The coordinate directions
$X^0,X^1,X^2,X^9$ are not compactified and are, henceforth, not
considered. $\Lambda$ is generated by the vectors $e_i \in \mathbb{R}^6$,
\begin{subequations}
  \begin{align}
    e_1 &= \alpha \left( 1, 0, 0, 0, 0, 0 \right), &
    e_2 &= \alpha \left( -\frac{1}{2}, \frac{1}{2}\sqrt{3}, 0, 0, 0, 0 \right) \\
    e_3 &= \alpha \left( 0, 0, 1, 0, 0, 0 \right), &
    e_4 &= \alpha \left( 0, 0, -\frac{1}{2}, \frac{1}{2}\sqrt{3}, 0, 0 \right) \\
    e_5 &= \alpha \left( 0, 0, 0, 0, 1, 0 \right), &
    e_6 &= \alpha \left( 0, 0, 0, 0, -\frac{1}{2}, \frac{1}{2}\sqrt{3} \right).
   \end{align}
  \label{eq:latticebasis}
\end{subequations}
The parameter $\alpha > 0$ is an arbitrary scale factor and is
adjusted, such that the winding modes are massive. Geometrically,
$2\pi\alpha$ is the circumference of the compactification torus.
The lattice factors
into three, mutually orthogonal rank $2$ lattices each being isomorphic
to an $\SU(3)$ root lattice. One could also assign different scale
factors to each of the factors. In the following the particular scale factors
are not important, as long as they do not give rise to massless
winding modes, so that I take them simply as equal.

The point group $P$ is generated by the elements $\tau$ and $\sigma$,
which act on $\Lambda$ in the lattice basis (\ref{eq:latticebasis}) by
  \begin{align}
    \sigma &= \left( \begin{array}{cccccc}
         0 & -1 &  0 &  0 &  0 &  0 \\
         1 & -1 &  0 &  0 &  0 &  0 \\
         0 &  0 & -1 &  1 &  0 &  0 \\
         0 &  0 & -1 &  0 &  0 &  0 \\
         0 &  0 &  0 &  0 &  1 &  0 \\
         0 &  0 &  0 &  0 &  0 &  1 \end{array} \right), &
     \tau &= \left( \begin{array}{cccccc}
         1 & -1 &  0 &  0 &  0 &  0 \\
         0 & -1 &  0 &  0 &  0 &  0 \\
         0 &  0 &  1 & -1 &  0 &  0 \\
         0 &  0 &  0 & -1 &  0 &  0 \\
         0 &  0 &  0 &  0 & -1 &  0 \\
         0 &  0 &  0 &  0 &  0 & -1 \end{array} \right).
  \label{eq:geospec}
  \end{align}
The generators $\tau$ and $\sigma$ have order $2$ and $3$, respectively, and
satisfy $(\tau\sigma)^2 = \iota$. According to the presentation
(\ref{eq:s3presentation}), the point group $P$ is isomorphic to $S_3$.
In the third torus $T^2$ the element $\sigma$ acts trivially and the
quotient $T^2/P$ is a $\mathbb{Z}_2$ orbifold. In the Euclidean basis,
$\tau$ is a reflection in the first and second torus,
\begin{subequations}
  \begin{align}
    x^4 &\mapsto -x^4 \mbox{ and } \\
    x^6 &\mapsto -x^6.
  \end{align}
\end{subequations}
Two reflections at orthogonal planes are equivalent to a rotation about
$\pi$ in the plane spanned by their normal vectors. The orthogonal
matrices $\sigma$ and $\tau$ can be rewritten in the form
(\ref{eq:pointgroupfermion}),
\begin{subequations}
\begin{align}
  \tau &\equiv \exp\left( 2\pi\imath \frac{1}{2} \left( J_{46} -
      J_{78} \right) \right)
  \label{eq:s3geotau}\\
  \sigma &\equiv \exp\left( 2\pi\imath \frac{1}{3} \left( J_{34} -
      J_{56} \right) \right).
  \label{eq:s3geosigma}
\end{align}
\end{subequations}
The twist vectors $\xi_g^{\mathrm{(F)}}$ and basis elements $D_i^g$ defining
the action in the antiholomorphic fermion sector for the representatives
$\iota$, $\tau$ and $\sigma$ of the point group conjugacy classes are given by
\begin{subequations}
  \begin{align}
    \xi_\iota^{\mathrm{(F)}} &= \left(0,0,0,0\right), & (D_i^\iota) &= \left(
      J_{34}, J_{56}, J_{78}, J_{12} \right) \\
    \xi_\tau^{\mathrm{(F)}} &= \left(\frac{1}{2},-\frac{1}{2},0,0\right), &
    (D_i^\tau) &= \left( J_{46}, J_{78}, J_{35}, J_{12} \right) \label{eq:s3taugeo}\\
    \xi_\sigma^{\mathrm{(F)}} &= \left(\frac{1}{3},-\frac{1}{3},0,0\right), &
    (D_i^\iota) &= \left( J_{34}, J_{56}, J_{78}, J_{12} \right) \label{eq:s3sigmageo}.
  \end{align}
  \label{eq:fermiontwistvectors}
\end{subequations}
In this model the point group
acts only on the first $16$ fermions by a \emph{standard
  embedding}. The point group is embedded into $\SO(16)$ as a subgroup
acting only on the first six fermions. Denote a basis for a Cartan
subalgebra of $\mathfrak{so}(16)$ by the matrices $K_{i,j}$
generating counterclockwise rotations in the
$\Xi^i$--$\Xi^j$--plane, i.e.~it is an antisymmetric $16\times 16$
matrix whose non-zero entries are $1$ in the $i$th row and $j$th
column and $-1$ in the $j$th row and $i$th column. The twist
vectors $\xi_g^{\mathrm{(YM)}}$ and the basis matrices $K_g^i$ are given by
\begin{subequations}
  \begin{align}
    \xi_\iota^{\mathrm{(YM)}} &= \left(0,0,0,0,0,0,0,0\right), \notag\\ (K_i^\iota) &= \left(
      K_{1,2}, K_{3,4}, K_{5,6}, K_{7,8}, K_{9,10}, K_{11,12},
      K_{13,14}, K_{15,16}\right) 
  \end{align} 
  \begin{align}
\xi_\tau^{\mathrm{(YM)}} &=
\left(\frac{1}{2},-\frac{1}{2},0,0,0,0,0,0\right), \notag\\ (K_i^\tau) &=
    \left( K_{2,4}, K_{5,6}, K_{1,3}, K_{7,8}, K_{9,10}, K_{11,12},
      K_{13,14}, K_{15,16}\right) \label{eq:s3tauym}
  \end{align} 
  \begin{align}
  \xi_\sigma^{\mathrm{(YM)}} &=
    \left(\frac{1}{3},-\frac{1}{3},0,0,0,0,0,0\right), \notag\\ (K_i^\sigma)
    &= \left( K_{1,2}, K_{3,4}, K_{5,6}, K_{7,8}, K_{9,10}, K_{11,12}.
      K_{13,14}, K_{15,16}\right). \label{eq:s3sigmaym}
 \end{align}
  \label{eq:ymtwistvectors}
\end{subequations}
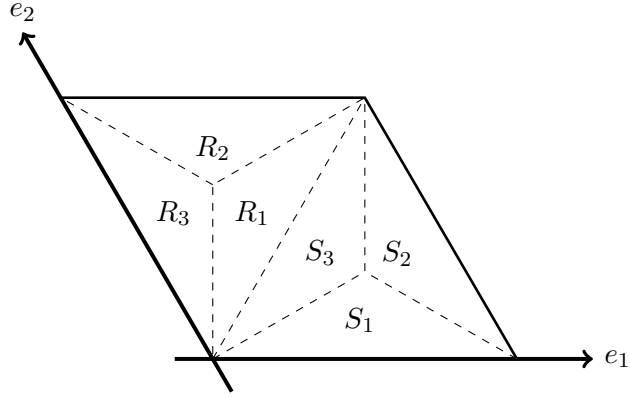
\begin{figure}
  \centering
    \begin{tikzpicture}
      \draw[line width=1pt] (0,0) -- (4,0) -- ++(120:4) -- ++(180:4)
      -- cycle;

      \draw[dashed] (0,0) -- (30:2.31) -- (60:4) -- (90:2.31) --
      cycle (4,0) -- (30:2.31) (90:2.31) -- (120:4);      

      \path (75:2) node {$R_1$};
      \path (90:2.8) node {$R_2$};
      \path (105:2) node {$R_3$};
      \path (45:2) node {$S_3$};
      \path (30:2.8) node {$S_2$};
      \path (15:2) node {$S_1$};

      \draw[dashed] (60:0) -- (60:4);
      \draw[line width=1.5pt,->] (0:-0.5) -- (0:5) node[anchor=west] {$e_1$};
      \draw[line width=1.5pt,->] (120:-0.5) -- (120:5) node[anchor=south] {$e_2$};
    \end{tikzpicture}

    \caption{Visualization of the actions of rotations about $2\pi/3$
      and a reflection at the $e_1$-axis on different regions $R_i$ and
      $S_i$ of the fundamental domain of the torus. The rotation maps
      $R_1 \rightarrow R_2 \rightarrow R_3 \rightarrow R_1$ and $S_1 \rightarrow S_2
      \rightarrow S_3 \rightarrow S_1$ and the reflection maps $S_1
      \leftrightarrow R_2$,
      $S_2 \leftrightarrow R_1$ and $S_3 \leftrightarrow R_3$.}
    \label{fig:s3orbifold}
\end{figure}
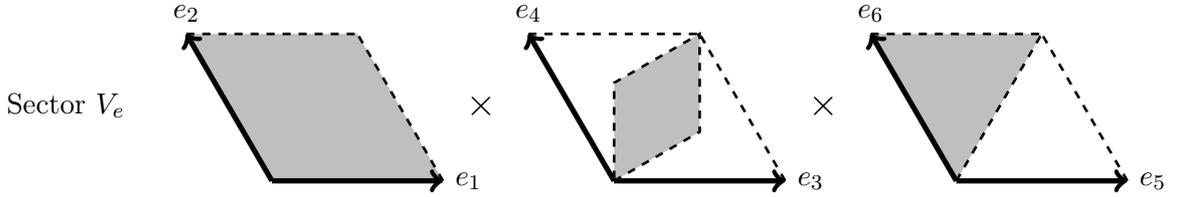
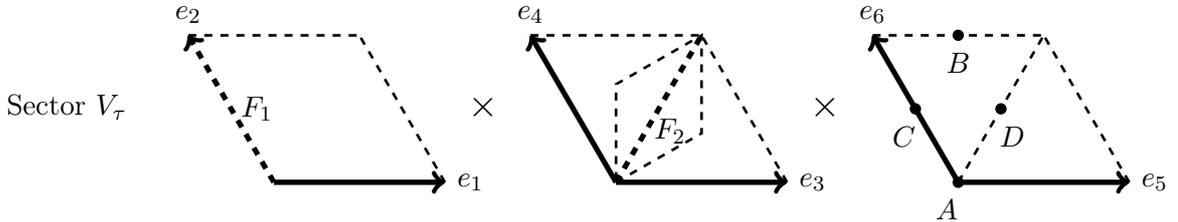
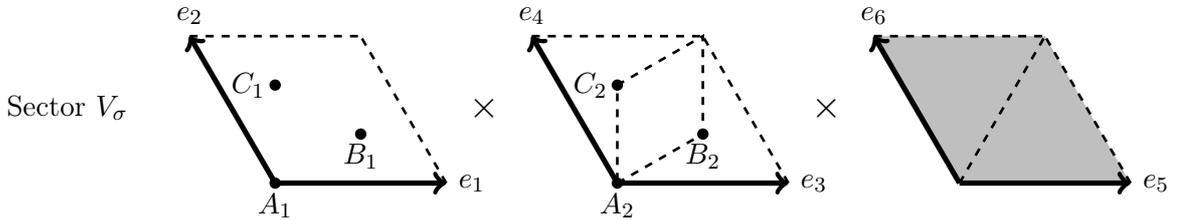
\begin{figure}[h]
  \centering
  \subfloat[A possible choice of fundamental domain (gray) for the $S_3$
  action on $T^6$. This is also the region where untwisted strings are
  localized.]{ 
  \begin{tikzpicture}
    \path (-1.8,1) node[anchor=east] {Sector $V_{\iota}$};

   \begin{scope}[cm={0.5,0,0,0.5,(0,0)}]  
      \fill[color=lightgray] (0:0) -- (0:4.5) -- (60:4.5) --
                       (120:4.5) -- cycle;
      \draw[line width=2pt,color=black,->] (0:0) -- (0:4.5)
      node[anchor=west] {$e_1$};
      \draw[line width=1pt,color=black,dashed] (0:4.5) -- (60:4.5);
      \draw[line width=2pt,color=black,->] (0:0) -- (120:4.5)
      node[anchor=south] {$e_2$};
      \draw[line width=1pt,color=black,dashed] (120:4.5) -- (60:4.5);   
    \end{scope} 
    
    \path (2.75,1) node[anchor=center] {\Large $\times$ };
    
    \begin{scope}[cm={0.5,0,0,0.5,(4.5,0)}]  
      \fill[color=lightgray] (0:0) -- (30:2.6) -- (60:4.5) --
                       (90:2.6) -- cycle;
      \draw[line width=1pt,dashed] (0:0) -- (30:2.6) -- (60:4.5) --
                       (90:2.6) -- cycle;
      \draw[line width=2pt,color=black,->] (0:0) -- (0:4.5)
      node[anchor=west] {$e_3$};
      \draw[line width=1pt,color=black,dashed] (0:4.5) -- (60:4.5);
      \draw[line width=2pt,color=black,->] (0:0) -- (120:4.5)
      node[anchor=south] {$e_4$};
      \draw[line width=1pt,color=black,dashed] (120:4.5) -- (60:4.5);   
    \end{scope} 
    
    \path (7.25,1) node[anchor=center] {\Large $\times$ };
 
    \begin{scope}[cm={0.5,0,0,0.5,(9,0)}]  

      \fill[color=lightgray] (0:0) -- (60:4.5) -- (120:4.5) -- cycle;
      \draw[line width=1pt,color=black,dashed] (0,0) -- (60:4.5);
            
      \draw[line width=2pt,color=black,->] (0:0) -- (0:4.5)
      node[anchor=west] {$e_5$};
      \draw[line width=1pt,color=black,dashed] (0:4.5) -- (60:4.5);
      \draw[line width=2pt,color=black,->] (0:0) -- (120:4.5)
      node[anchor=south] {$e_6$};
      \draw[line width=1pt,color=black,dashed] (120:4.5) -- (60:4.5);
    \end{scope} 
  \end{tikzpicture}}

  \subfloat[In the $\tau$--twisted sector there are $4$
  fixed points. The string is localized on a torus $F_1 \times
  F_2$. $F_i$ are fixed circles and shown as bold dashed lines.]{
  \begin{tikzpicture}
    \path (-1.8,1) node[anchor=east] {Sector $V_\tau$};

   \begin{scope}[cm={0.5,0,0,0.5,(0,0)}]  
      \draw[line width=2pt,color=black,->] (0:0) -- (0:4.5)
      node[anchor=west] {$e_1$};
      \draw[line width=1pt,color=black,dashed] (0:4.5) -- (60:4.5);
      \draw[line width=1pt,color=black,dashed] (120:4.5) -- (60:4.5);  
      \draw[line width=2pt,color=black,dashed,->] (0:0) -- (120:4.5)
      node[anchor=south] {$e_2$};

      \path (120:2.25) node[anchor=west] {$F_1$};
    \end{scope} 
    
    \path (2.75,1) node[anchor=center] {\Large $\times$ };

    \begin{scope}[cm={0.5,0,0,0.5,(4.5,0)}] 
      \draw[line width=1pt,dashed] (0:0) -- (30:2.6) -- (60:4.5) --
                       (90:2.6) -- cycle;
      \draw[line width=2pt,color=black,->] (0:0) -- (0:4.5)
      node[anchor=west] {$e_3$};
      \draw[line width=1pt,color=black,dashed] (0:4.5) -- (60:4.5);
      \draw[line width=2pt,color=black,->] (0:0) -- (120:4.5)
      node[anchor=south] {$e_4$};
      \draw[line width=1pt,color=black,dashed] (120:4.5) -- (60:4.5);   
 
      \draw[line width=2pt,color=black,dashed] (0:0) -- (60:4.5);

      \path (60:2.25)+(0.2:0.3) node[anchor=north] {$F_2$};
    \end{scope} 
        
    \path (7.25,1) node[anchor=center] {\Large $\times$ };
 
    \begin{scope}[cm={0.5,0,0,0.5,(9,0)}]  
      \draw[line width=1pt,color=black,dashed] (0,0) -- (60:4.5);
            
      \draw[line width=2pt,color=black,->] (0:0) -- (0:4.5)
      node[anchor=west] {$e_5$};
      \draw[line width=1pt,color=black,dashed] (0:4.5) -- (60:4.5);
      \draw[line width=2pt,color=black,->] (0:0) -- (120:4.5)
      node[anchor=south] {$e_6$};
      \draw[line width=1pt,color=black,dashed] (120:4.5) -- (60:4.5); 
      
      \fill[color=black] (0:0) circle(1.5mm) (90:3.9) circle(1.5mm)
      (120:2.25) circle(1.5mm) (60:2.25) circle(1.5mm);
      
      \path (0:0)+(-0.3,-0.2) node[anchor=north] {$A$};
      \path (90:3.7) node[anchor=north] {$B$};
      \path (120:2.25)+(-0.3,-0.2) node[anchor=north] {$C$};
      \path (60:2.25)+(0.3,-0.2) node[anchor=north] {$D$};
    \end{scope}  
  \end{tikzpicture}
}

\subfloat[In the $\sigma$--twisted sector there are $9$ fixed points
$(A_1,A_2)$, etc.. In the third torus the string is localized in the
bulk.]{ 
  \begin{tikzpicture}
    \path (-1.8,1) node[anchor=east] {Sector $V_\sigma$};

   \begin{scope}[cm={0.5,0,0,0.5,(0,0)}]  
      \draw[line width=2pt,color=black,->] (0:0) -- (0:4.5)
      node[anchor=west] {$e_1$};
      \draw[line width=1pt,color=black,dashed] (0:4.5) -- (60:4.5);
      \draw[line width=2pt,color=black,->] (0:0) -- (120:4.5)
      node[anchor=south] {$e_2$};
      \draw[line width=1pt,color=black,dashed] (120:4.5) -- (60:4.5);   

      \fill[color=black] (0:0) circle(1.5mm) (30:2.6) circle(1.5mm)
                       (90:2.6) circle (1.5mm);
      \path (0:0) node[anchor=north] {$A_1$};
      \path (30:2.6) node[anchor=north] {$B_1$};
      \path (90:2.6) node[anchor=east] {$C_1$};
    \end{scope} 
    
    \path (2.75,1) node[anchor=center] {\Large $\times$ };
    
    \begin{scope}[cm={0.5,0,0,0.5,(4.5,0)}]  
      \draw[line width=1pt,dashed] (0:0) -- (30:2.6) -- (60:4.5) --
                       (90:2.6) -- cycle;
      \draw[line width=2pt,color=black,->] (0:0) -- (0:4.5)
      node[anchor=west] {$e_3$};
      \draw[line width=1pt,color=black,dashed] (0:4.5) -- (60:4.5);
      \draw[line width=2pt,color=black,->] (0:0) -- (120:4.5)
      node[anchor=south] {$e_4$};
      \draw[line width=1pt,color=black,dashed] (120:4.5) -- (60:4.5);   

      \fill[color=black] (0:0) circle(1.5mm) (30:2.6) circle(1.5mm)
                       (90:2.6) circle (1.5mm);
      \path (0:0) node[anchor=north] {$A_2$};
      \path (30:2.6) node[anchor=north] {$B_2$};
      \path (90:2.6) node[anchor=east] {$C_2$};
    \end{scope} 
    
    \path (7.25,1) node[anchor=center] {\Large $\times$ };
 
    \begin{scope}[cm={0.5,0,0,0.5,(9,0)}]  
      \fill[color=lightgray] (0:0) -- (0:4.5) -- (60:4.5) -- (120:4.5) -- cycle;
      \draw[line width=1pt,color=black,dashed] (0,0) -- (60:4.5);
            
      \draw[line width=2pt,color=black,->] (0:0) -- (0:4.5)
      node[anchor=west] {$e_5$};
      \draw[line width=1pt,color=black,dashed] (0:4.5) -- (60:4.5);
      \draw[line width=2pt,color=black,->] (0:0) -- (120:4.5)
      node[anchor=south] {$e_6$};
      \draw[line width=1pt,color=black,dashed] (120:4.5) -- (60:4.5); 
    \end{scope} 
  \end{tikzpicture}}

  \caption{Fundamental domain and localization of strings in the $T^6/S_3$ orbifold.}
  \label{fig:s3sectors}
\end{figure}
The point group has three conjugacy
classes and so there are three twisted sectors, labeled by the elements
$\iota$, $\tau$ and $\sigma$.
The geometry of this orbifold might be unfamiliar to the reader. In
order to find a fundamental domain for the action of $S_3$ on the
torus $T^6$ it is helpful to first get used to the action of $\tau$
and $\sigma$ acting on a two dimensional torus. In figure
\ref{fig:s3orbifold} the fundamental region of the torus has been
divided into six subdomains. The action of a counterclockwise rotation
about $2\pi/3$ and a reflection at the $e_1$-axis are described in its
caption. 
Divide now the torus $T^6$ into subdomains of the form $D \times D'
\times D''$ where $D$, $D'$ and $D''$ are one of the subdomains $R_i$
or $S_i$ from figure \ref{fig:s3orbifold}. One obtains this way $6^3 =
216$ subdomains of $T^6$. Using the definition of $\sigma$ and $\tau$,
one can find exactly $6^2 = 36$ of those subdomains that are not
identified under the action of $S_3$. The union of those subdomains
constitutes a fundamental domain. A possible choice of fundamental
domain is shown in figure \ref{fig:s3sectors}. 
Figure \ref{fig:s3sectors} also shows the regions of $T^6/S_3$ in
which the twisted strings are localized. From there it can be seen
that the effective geometry for the $\sigma$ and $\tau$ twisted
sectors is six dimensional, while it is ten dimensional for the
untwisted string.

In the $\sigma$ twisted sector there are nine fixed-tori
$T^\sigma_{(P,Q)}$. Each of the tori can be labelled by a pair $(P,Q)$
where $P \in \lbrace A_1,B_1,C_1 \rbrace$ is a fixed point in the
first torus and $Q \in \lbrace A_2,B_2,C_2\rbrace$ is one in the
second torus. Note that $\tau$ interchanges $C_1 \leftrightarrow B_1$
and $C_2 \leftrightarrow B_2$. Nevertheless this does not identify
some of the fixed-tori, because $\tau$ does not commute with $\sigma$
and so changes boundary conditions to $\sigma^2$. There are no
restrictions on the momentum in the third torus.
In the $\tau$ twisted sector four fixed-tori can be found. Each of
these tori intersect the torus $T^\sigma_{(A_1,A_2)}$ in exactly one
point. All other fixed-tori are mutually disjoint.

\subsection{The physical spectrum}
This $S_3$ model possesses four dimensional $N=1$ SUSY and an
$\E_6\times\U(1)$ Yang--Mills symmetry. Additionally, there is an $R$
symmetry that contains $\mathbb{Z}_{4}^R$ and a $\mathbb{Z}_2$
parity. The spectrum can be found in table \ref{tab:s3modelspectrum}
in the appendix.

\subsubsection{Orbifolded fermion sector}
\label{sec:fermionsector}
The fermionic sector consists of eight antiholomorphic fermions
$\psi^i(\bar{z})$ and can be analyzed by the method described in
section \ref{sec:orbifolds-strings-projection} with the shift vectors and
Cartan subalgebras given in equation
(\ref{eq:fermiontwistvectors}). States are classified according to
irreducible representations of
the subalgebra $\mathfrak{g}_0$ of $\mathfrak{so}(8)$ that is
invariant under the action of the point group. In the case at hand,
the breaking scheme is
\begin{align}
  \mathfrak{so}(8) &\rightarrow \mathfrak{u}(1) \oplus \mathfrak{u}(1)
  \oplus \mathfrak{u}(1) = \mathfrak{g}_{0}.
\end{align}

The generators of the various invariant $\mathfrak{u}(1)$ algebras are
given by
\begin{subequations}
  \begin{align}
    R_0 &= J_{12}, \\
    R_1 &= J_{35} - J_{46} \\
    R_2 &= J_{78}.
  \end{align}
\end{subequations}
The charge $R_0$ generates rotations in the uncompactified directions
$x^1$ and $x^2$ and corresponds to the \emph{helicity operator}. The
charges $R_1$ and $R_2$ generate rotations in the
compactified directions and must respect the lattice structure of
$\Lambda$. As lattice automorphism groups are always finite, $R_1$ and
$R_2$ generate a discrete symmetry. Call the eigenvalues of $2 R_1$
and $2 R_2$ $D$ charges ($D$ for discrete). Table
\ref{tab:fermionsector} gives the states in the fermion sector with
conformal weight $\leq 1/2$. 

\paragraph{Discrete symmetries and $R$ symmetry}
The charge $R_2$ generates rotations in the $x^7-x^8$-plane. In order
to be compatible with the orbifold geometry, these rotations have to
be lattice automorphisms, cf. section
\ref{sec:orbifolds-strings-projection}. In the $x^7-x^8$-plane the torus is
defined by an $\SU(3)$ root lattice. It is a hexagonal lattice and allows only
for rotations about $2\pi/6$. Hence, there is a
$\mathbb{Z}_{12}$ symmetry generated by
\begin{align}
  \beta &= \exp\left( 2\pi\imath \frac{1}{12} \cdot 2J_{78} \right).
\end{align}
The order of the generator is larger than $6$, because $J_{78}$ can
have half--integer eigenvalues. In the orbifold model, the operators
$U(\beta)$ map states localized at one fixed point to another, see
figure \ref{fig:dsymmorbi}. Compatibility with the orbifold geometry
requires that only $\beta^3$ is a symmetry of the full model and the
discrete symmetry is reduced to $\mathbb{Z}_4$.
\begin{figure}
  \centering
  \subfloat[The rotation $\beta$ exchanges the fixed points
  on the third torus: $B\rightarrow C\rightarrow D\rightarrow B$]{
    \begin{tikzpicture}
    \begin{scope}[cm={0.8,0,0,0.8,(0,0)}]  
      \draw[line width=1pt,color=black,dashed] (0:0) -- (60:4.5);
            
      \draw[line width=2pt,color=black,->] (0:0) -- (0:4.5)
      node[anchor=west] {$e_5$};
      \draw[line width=1pt,color=black,dashed] (0:4.5) -- (60:4.5);
      \draw[line width=2pt,color=black,->] (0:0) -- (120:4.5)
      node[anchor=south] {$e_6$};
      \draw[line width=1pt,color=black,dashed] (120:4.5) -- (60:4.5);
      
      \fill[color=black] (0:0) circle(1.5mm) (90:3.9) circle(1.5mm)
      (120:2.25) circle(1.5mm) (60:2.25) circle(1.5mm);
      
      \path (0:0)+(-0.3,-0.2) node[anchor=north] {$A$};
      \path (90:3.9)+(0.3,0.2) node[anchor=south] {$B$};
      \path (120:2.25)+(-0.3,-0.2) node[anchor=north] {$D$};
      \path (60:2.25)+(0.3,-0.2) node[anchor=west] {$C$};

      \draw[line width=2pt,<-] (120:2.25)+(5:2.25) arc(5:55:2.25);
      \draw[line width=2pt,->] (90:3.9)+(-65:2.25) arc(-65:-115:2.25);
      \draw[line width=2pt,->] (60:2.25)+(175:2.25) arc(175:125:2.25);
    \end{scope}  
    \end{tikzpicture}
}

\subfloat[The rotation $\alpha$ permutes the $9$ fixed points
of the $\sigma$--twisted sector.]{
  \begin{tabular}{ccc}
    $(A_1,A_2)$ &$\rightarrow$& $(A_1,A_2)$ \\
    $(A_1,B_2)$ &$\rightarrow$& $(B_1,A_2)$ \\
    $(A_1,C_2)$ &$\rightarrow$& $(C_1,A_2)$
  \end{tabular}
  \begin{tabular}{ccc}
    $(B_1,A_2)$ &$\rightarrow$& $(A_1,C_2)$ \\
    $(B_1,B_2)$ &$\rightarrow$& $(B_1,C_2)$ \\
    $(B_1,C_2)$ &$\rightarrow$& $(C_1,C_2)$
  \end{tabular}
  \begin{tabular}{ccc}
    $(C_1,A_2)$ &$\rightarrow$& $(A_1,B_2)$ \\
    $(C_1,B_2)$ &$\rightarrow$& $(B_1,B_2)$ \\
    $(C_1,C_2)$ &$\rightarrow$& $(C_1,B_2)$ 
  \end{tabular}
}

  \caption{The action of the rotations $\alpha$ and $\beta$
    on the localization of the states. Only $\beta^3$ and $\alpha^4$
    are compatible with the orbifold geometry. For $\alpha$ the
    different localization points of winding modes in the
    $\tau$--sector have to be taken into account. The labels for the
    fixed-points are the same as in figure \ref{fig:s3sectors}.}
  \label{fig:dsymmorbi}
\end{figure}
Similarly, $R_1$ generates rotations in
the $x^3-x^5$--plane together with an opposite rotation in the
$x^4-x^6$--plane by the same angle. The lattice $\Lambda$ defining the
torus $T^6$ in these planes is a square lattice. Therefore, rotation angles 
have to be restricted to multiples of $2\pi/4$. Consequently, there exists
a $\mathbb{Z}_8$ symmetry generated by
\begin{align}
  \alpha &= \exp\left( 2\pi\imath \frac{1}{8} \cdot 2(J_{35}-J_{46}) \right).
\end{align}
$U(\alpha)$ permutes the fixed points of the $\sigma$-twisted
sector, cf. figure \ref{fig:dsymmorbi}. Thus, only $\alpha^4$ is
compatible with the orbifold geometry and the symmetry reduces to a
$\mathbb{Z}_2$ parity. 

The charges under $\mathbb{Z}_8 \times \mathbb{Z}_{12}$ are shown
in table \ref{tab:fermionsector} as ``$D$ charges''. From the spectrum
it can be deduced that the supercharge in four dimensions must be
charged under $\mathbb{Z}_2 \times \mathbb{Z}_4$ with charge $(0,1)$
and $\beta^3$ generates an $R$ symmetry. The generator $\alpha^4$
commutes with the supercharge and is a discrete non--$R$ symmetry. The
$R$ and $\mathbb{Z}_2$ charges of the particles are listed in table
\ref{tab:s3modelspectrum}.

\subsubsection{Orbifolded Yang--Mills Sector}
In the Yang--Mills sector the
invariant subalgebra $\mathfrak{g}_{0}$ of $\mathfrak{so}(16)$ is
obtained from the symmetry breaking
\begin{subequations}
\begin{align}
  \mathfrak{so}(16) &\rightarrow \mathfrak{so}(10) \oplus
  \mathfrak{u}(1) \oplus \mathfrak{u}(1) = \mathfrak{g}_{0}.
\end{align}
In the fermionic realization the AKM symmetry is enhanced. The
enhanced symmetry breaking pattern is
\begin{align}
  \mathfrak{e}_8 &\rightarrow \mathfrak{e}_6 \oplus \mathfrak{u}(1).
\end{align}
The rank of the symmetry algebra has been reduced. This
is one of the features that distinguishes Abelian orbifold models
from non--Abelian orbifold models.
\end{subequations}
Table \ref{tab:ymsector} shows the states in this sector with
conformal weight $\leq 1$. The $\U(1)$ charge $Q$ of $\E_6
\times \U(1)$ is related to the two $\U(1)$ charges, $A_1$ and $A_2$, of
$\SO(10)\times \U(1) \times \U(1)$ by
\begin{align}
  Q &= A_1 + 2 A_2.
\end{align}

\subsubsection{Massless states}
The full string state is a vector in the ghost extended state
space. Because both the world--sheet super current and the
stress--energy tensor are invariant under orbifolding, the conformal
and superconformal ghost sectors need not to be changed. An on--shell
string state has to satisfy the equations of motion
(\ref{eq:hstmass2}) and has to be invariant under the action of the
point group $P$. Using intermediate results from tables
\ref{tab:fermionsector} and \ref{tab:ymsector} it is possible to
recover the full massless spectrum from table \ref{tab:s3modelspectrum}.

\subsection{Discussion}
The $T^6/S_3$ orbifold model described in this section exhibits a
non-chiral spectrum in four dimensions. Therefore, its gauge
symmetries are automatically free of chiral anomalies. This is an
important requirement for consistency of the model
\cite{Fujikawa,Lee,Green,Tosa,Chen}.

From figure \ref{fig:s3sectors} it can be seen that the $\tau$- and
$\sigma$-twisted sectors are indeed six dimensional, because both have
fixed-tori. From table \ref{tab:fermionsector} one sees that in the
$\tau$-twisted sectors the states comprise a 4D, $\mathcal{N}=1$
chiral supermultiplet. The effective six-dimensional theory on the
fixed-tori is $\mathcal{N}=(1,0)$ supersymmetric and is localized in
the $X^1$, $X^2$, $X^3$, $X^5$-transversal directions. Two chiral
multiplets then organize into a hypermultiplet. The intermediate gauge
group is $\E_7\times\SU(2)$ and the hypermultiplet transforms as
$(\mathbf{56},\mathbf{1})$. Note that on every fixed-torus one finds
only $28$ hypermultiplets. But, because $\mathbf{56}$ is a pseudoreal
representation of $\E_7$, $28$ hypermultiplets are sufficient to
realize one half-hypermultiplet transforming as $(\mathbf{56},\mathbf{1})$
\cite{Schwarz:1995zw,Witten:1995gx}. Similarly, one finds at each
fixed-torus $4$ half-hypermultiplets transforming as
$(\mathbf{1},\mathbf{2})$ built up 
from one half-hypermultiplet each, as $\mathbf{2}$ is also pseudoreal. As
these states involve holomorphic oscillators, they can be identified
with $\tau$-twisted moduli of this orbifold.

A similar analysis can be applied to the $\sigma$-twisted sector. As
before one finds that the effective six-dimensional theory at each of
the $9$ fixed-tori is $\mathcal{N}=(1,0)$ supersymmetric and that the
fields are localized in the $X^1$, $X^2$, $X^7$, $X^8$-transversal direction. The
intermediate gauge group is $\E_7\times\U(1)$ in this case. Matter
organizes into one half-hypermultiplet transforming as $\mathbf{56}_0$ and
$7$ singlet half-hypermultiplets. The latter involve again oscillators and
are therefore identified with $\sigma$-twisted moduli.

Note that the above $g$-twisted states are only invariant under the
centralizer of $g$ in the point group. This means that the twisted
sectors can be almost considered as the twisted sectors of a
$\mathbb{Z}_2$- or $\mathbb{Z}_3$-orbifold. The spectrum agrees almost with
the one given in table $2$ in \cite{Aldazabal:1997wi}. One has to take
into account that in the $S_3$-orbifold the number of fixed-tori and
twisted sectors is different from the Abelian orbifolds considered
there. For example, the $\mathbb{Z}_3 = \langle \sigma \rangle$
orbifold model has two twisted sectors corresponding to the elements
$\sigma$ and $\sigma^2$. In the $S_3$ orbifold model both elements are
related by conjugation by $\tau$ and are therefore identified. Since
in the $\sigma$-twisted sectors the number of fixed-tori is the same
in both models, the only difference that has to be taken into account
is that instead of having matter particles transforming as
half-hypermultiplets, the matter particles transfrom as
half-hypermultiplets. Conversely, in the $\tau$-twisted sector the
number of fixed-tori is reduced from $16$ in \cite{Aldazabal:1997wi}
to $4$ in the $S_3$-orbifold. Hence, one ends up with only one quarter
of the matter contents of \cite{Aldazabal:1997wi}.

Since the twisted matter is organized into half-hypermultiplets,
dimensional reduction to four dimensions automatically yields a
non-chiral spectrum. This means that this particular orbifold model
is not phenomenologically viable. It is not possible to obtain a
chiral theory from this orbifold without introducing non-trivial
background fields \cite{Fischer:2013qza}.

\subsection[{Untwisted moduli of the $S_3$ orbifold}]{Untwisted moduli
  of the {\boldmath $S_3$} orbifold}
\label{sec:quantumgeometry}
Connected to given a state in string theory there is associated a
family of infinitesimally deformations. Introducing complex
coordinates allows one to distinguish between complex structure moduli
and K\"ahler moduli. A possible choice of complex coordinates with
this property is given by
\begin{align}
  z^1 = x^3 + \imath x^5,\;
  z^2 = x^6 + \imath x^4,\;
  z^3 = x^7 + \imath x^8.  
  \label{eq:ccordinates}
\end{align}

In the coordinates $z^i$ the maps $\tau$ and $\sigma$ act
holomorphically on the torus $T^6$. Their action on a tuple
$(z_1,z_2,z_3)$ can be deduced from equation (\ref{eq:geospec}) and
reads as
\begin{subequations}
  \begin{align}
    (z_1,z_2,z_3) &\overset{\tau}{\rightarrow} (z_1,-z_2,-z_3) \\
    (z_1,z_2,z_3) &\overset{\sigma}{\rightarrow} (z_1
    \cos\frac{2\pi}{3} - \imath z_2 \sin\frac{2\pi}{3},-\imath z_1
    \sin\frac{2\pi}{3} + z_2 \cos\frac{2\pi}{3},z_3).
  \end{align}
  \label{eq:holaction}
\end{subequations}
In complex coordinates $\delta H_{ab}$ and $\delta H_{a\bar{b}}$ are
the deformations of the complex structure and K\"ahler structure,
respectively. Massless deformations are given by
  \begin{align}
    \delta H_{1\bar{1}} + \delta H_{2\bar{2}},&\;\;\delta H_{11} -
    \delta H_{22},\;\;
    \delta H_{3\bar{3}},\;\;\delta H_{33}
    \label{eq:moduli}
  \end{align}
together with their complex conjugates. Hence, there are two complex
structure moduli and two K\"ahler moduli. These numbers agree with the
numbers given in table \ref{tab:s3modelspectrum} as well as with the number
of global $(2,1)$ and $(1,1)$ forms on the orbifold as can be checked
explicitly using the holomorphic action (\ref{eq:holaction}).

\section{Conclusions}
\label{sec:conclude}
In this paper I constructed a heterotic orbifold model with non
Abelian point group $S_3$. However, the methods presented here apply
to more general toroidal symmetric orbifolds as well. The only
restriction is that the left moving Yang--Mills sector and right
moving fermion sector are not mixed by orbifolding. The Yang--Mills
sectors is realized by a free fermion CFT, so that the analysis of the
state space is formally the same as for the right moving fermion
sector. In this model the YM gauge group is broken from $\E_8 \times \E_8$
down to $\E_6\times\U(1)\times\E_8$, i.e.~it reduces the rank by
one. Rank reduction on the orbifold itself cannot be achieved by using Abelian point
groups. Hitherto, it could only be achieved by using a non--diagonal
embedding of the space group or in the
low energy effective action via some Higgs mechanism.
Compared to Abelian orbifold model building the only new technical
points are the appearance of multidimensional representations of the
point group. Additionally, the method used here to analyze the
spectrum requires one to change the Cartan subalgebra for each
twisted sector. The first point is handled by using the representation
theory of finite groups, while the second point is more subtle. The
fermionic realization treats all Yang--Mills degrees of freedom
symmetrically and allows so for freely regrouping of the real fermions
into complex fermions. The drawback is here that the space group
action has to be embedded into $\SO(16)\times\SO(16)$. All in all the
methods presented here complement the well-known bosonic framework and
could be used to extend the search for phenomenologically realistic
string models to include non--Abelian point groups as well. However,
most different point groups have to be used. 

{\bf Acknowledgements.} I would like to thank Michael Ratz for useful
comments on the manuscript, giving
me the opportunity to join his research group and work on such an
interesting topic. Moreover, I would like to thank Jes\'us
Torrado--Cacho for reading the manuscript. Likewise, I want to express
my gratitude to all members of the T30e group at TUM for very
interesting and stimulating discussions.

\newpage
\appendix
\section{Fermionic realization of AKM symmetries and bosonization}
\label{app:fermionicakm}
The fermionic realization of the $\E_8\times\E_8$ affine symmetry
and its relation to the usual bosonized description play an important
role in my analysis of the orbifolding process. For the reader's
convenience I provide a short overview on this subject. It is pretty
standard and I refer the reader to any good textbook on conformal
field theory for more details.

\subsection{Twisted complex fermions and their bosonization}
\label{sec:bosonized}
\label{sec:fermionorbifold}
The CFT of $n$ real Majorana--Weyl fermions $\psi^i(z)$,
$i=1,\ldots,n$ is defined by the OPE and stress tensor $T(z)$ given by
\begin{subequations}
  \begin{align}
    \psi^i(z) \psi^j(w) &\sim \frac{\delta^{ij}}{z-w} \\
    \label{eq:fermionalg}
    T(z) &= -\frac{1}{2} \sum_{i=1}^n : \psi^i(z)\partial\psi^i(z):.
  \end{align} 
\end{subequations}
For two fermions one can define complex fermions $\Psi(z) =
\frac{1}{\sqrt{2}} \left( \psi^1(z) + \mathrm{i}\psi^2(z) \right)$ and
the fermion number current $j(z) = \mathrm{i}
:\Psi(z)\bar{\Psi}(z):$. W.r.t. the $\U(1)$ current $j(z)$ the field
$\Psi(z)$ has charge $+1$ and $\bar{\Psi}(z)$ has charge $-1$. It
turns out that the fermion field operators can be expressed as
composite operators of a single free boson $\phi(z)$ on the same world
sheet as
\begin{align}
  \Psi(z) = :\exp\left(\mathrm{i}\phi(z)\right):,\;\;
  \bar{\Psi}(z) = :\exp\left(-\mathrm{i}\phi(z)\right):,\;\;
  j(z) &= \imath\partial\phi(z).
  \label{eq:bosonizedfcurrent}
\end{align}
These identifications are called bosonization rules. The irreducible
representations of the bosonic field algebra are characterized by
momenta $p$. For bosonization to work, the boson has to be
compactified on a circle of radius $1$, i.e. the momenta $p$ in the NS
sector have to be integral $p\in\mathbb{Z}$, while the momenta in the
R sector fulfill $p\in\frac{1}{2}+\mathbb{Z}$ to account for the
antiperiodic boundary conditions. Moreover, the spin field $\sigma(z)$
is given by $:\exp\left(\frac{\mathrm{i}}{2}\phi(z)\right):$, i.e. it
changes the boundary conditions from periodic to antiperiodic and vice
versa. It is important to notice that the fermion number, i.e. the
zero mode of $j(z)$, is identical to the momentum. It is clear that
this construction can be generalized to arbitrary boundary conditions
obeyed by the complex fermions. The most generic boundary conditions
compatible with the form of the stress tensor are characterized by a
twist parameter $0 \leq s < 1$ and take the form
\begin{align}
  \Psi\left(z e^{2\pi\mathrm{i}} \right) &= -e^{- 2\pi\mathrm{i}
    s}\Psi\left(z\right).
\end{align}
The choice $s=0$ corresponds to the R sector and $s=\frac{1}{2}$ to
the NS sector. In the bosonized description the momenta take values in
a shifted lattice $p \in s-\frac{1}{2}+\mathbb{Z}$.

A generic state space is either specified by a certain number of
fermionic oscillators acting on a ground state or, equivalently, by
giving the ground state momentum $p$ and the number $N_{osc}$ of
bosonic oscillators acting on $|p\rangle$. The conformal weight $L_0$
of such a state is then given by 
\begin{align}
  L_0 &= \frac{1}{2} p^2 + N_{osc} + \delta c.
\end{align}
The number $\delta c$ is equal to the conformal weight of the ground
state, i.e.
\begin{align}
\delta c = \frac{1}{2} \left( s - \frac{1}{2}
\right)^2.
\label{eq:fermionicshift}
\end{align}

\subsection{Bosonized description and enhanced Yang--Mills symmetry}
\label{sec:enhanced}
The bosonization representation from section \ref{sec:bosonized} can be
used to describe the fermionic realization of $\SO(8)$ and
$\SO(16)$ symmetry in the heterotic string. For definiteness only the
holomorphic realization of $\SO(16)$ in terms of 16 Majorana--Weyl fermions
$\Xi^I(z)$, $I = 1,\ldots,16$, is discussed. The antiholomorphic
realization of $\SO(8)$ goes along the same lines. In view
of orbifolds, generalized twisted boundary conditions are included in
the discussion.

The $16$ Majorana--Weyl fermions $\Xi^I(z)$ can be grouped into $8$ complex
fermions $\Psi^i(z)$,
\begin{subequations}
  \begin{align}
    \Psi^i(z) &= \frac{1}{\sqrt{2}} \left( \Xi^{2i-1}(z) + \imath \Xi^{2i}(z)
    \right) \\
    \bar{\Psi}^i(z) &= \frac{1}{\sqrt{2}} \left( \Xi^{2i-1}(z) -
      \imath \Xi^{2i}(z) \right).
  \end{align}
  \label{eq:cfermion2}
\end{subequations}
The complex fermions (\ref{eq:cfermion2}) should obey twisted boundary
conditions. In the NS--sector the twist parameters $s^i$ are therefore
given by
\begin{subequations}
  \begin{align}
    s^i &\equiv \frac{1}{2} + \xi^i\mbox{ mod }1,
    \label{eq:twistns}
  \end{align}
while in the R--sector they read as
  \begin{align}
    s^i &\equiv \xi^i\mbox{ mod }1.
    \label{eq:twistr}
  \end{align}
\end{subequations}
By section \ref{sec:bosonized}, each complex fermion $\Psi^i(z)$ can be
represented by a free boson $\phi^i(z)$, i.e.
\begin{align}
  \Psi^i(z) = : \exp\left( \imath \phi^i(z) \right):,&\;
  \bar{\Psi}^i(z) = : \exp\left( -\imath \phi^i(z) \right):.
\end{align}
The momentum $p^i$ of the boson $\phi^i$ takes values in a
one--dimensional lattice,
\begin{align}
  p^i &\in s^i - \frac{1}{2} + \mathbb{Z}.
  \label{eq:Xshiftedmomentum}
\end{align}
A Cartan subalgebra of $\mathfrak{so}(16)$ is spanned by the
zero--modes of the currents $H^i(z) \equiv K^{2i-1,2i}(z)$. In terms
of the complex fermions $\Psi^i(z)$, they read as
\begin{align}
  H^i(z) &=  K^{2i-1,2i}(z) = \imath : \Xi^{2i-1}(z) \Xi^{2i}(z) : =
  j^i(z),
  \label{eq:fcsacurrent}
\end{align}
where $j^i(z)$ is the fermion number current. Therefore, the Cartan
generators $H^i_0$ of $\SO(16)$ correspond to fermion number operator
$F^i$ for the complex fermions $\Psi^i(z)$. Now, by equation
(\ref{eq:bosonizedfcurrent}) the fermion number current $j^i(z)$ can
be expressed in terms of the free bosons $\phi^i(z)$ as
\begin{align}
  \imath \partial \phi^i(z) &= j^i(z) = H^i(z),
\end{align}
which identifies the eigenvalues of the Cartan generators $H_0^i$ with
the momentum (\ref{eq:Xshiftedmomentum}).

The GSO projection selects states according to their $G$ parity, which
is determined by their fermion number, i.e. their momentum in the
bosonized description. The $G$ parity operator for a system of $8$
complex fermions is given by
\begin{align}
  G &= \prod_{i=1}^8 (-1)^{F^i + \frac{1}{2} - s^i}.
\end{align}
The shift of the fermion number by $\frac{1}{2} - s^i$ takes into
account that the ground states are assigned positive $G$ parity. The
$G$ parity operator intertwines between the different fermion
sectors. A momentum $(p^1,\ldots,p^8)$ is not projected out by the
GSO projection if
\begin{align}
  \sum_{i=1}^8 p^i + \frac{1}{2} - s^i \in 2\mathbb{Z}.
\end{align}
Therefore, in the GSO projected state space of the sector twisted by
$s$ the momenta are given by
\begin{align}
  p^i &= s^i - \frac{1}{2} + n^i,\;\;\sum_{i=1}^8 n^i \in 2\mathbb{Z}.
\end{align}
Consider the untwisted case, i.e. $\xi = 0$. In the NS--sector it is
$s^i = \frac{1}{2}$ and the momenta $(p^i)$ are restricted to values
whose sum is even. This lattice coincides with the root lattice
$\Lambda_{\SO(16)}$ of the Lie group $\SO(16)$. In the R--sector all
momenta are shifted by $\lambda_s = \left(
  \frac{1}{2},\ldots,\frac{1}{2} \right)$, which corresponds to the
highest weight vector of a spinor representation of $\SO(16)$. In the
twisted case, i.e. $\xi \neq 0$, the momenta are shifted by the vector
$(\xi^i)$,
\begin{subequations}
  \begin{align}
    (p^i) &\in \xi + \Lambda_{\SO(16)}\mbox{ in the NS--sector} \\
    (p^i) &\in \xi + \lambda_s + \Lambda_{\SO(16)}\mbox{ in the R--sector.}
  \end{align}
  \label{eq:ymmomenta}
\end{subequations}
In the bosonized description, the fermionic and bosonic stress energy 
tensors coincide, by section \ref{sec:bosonized}. In
particular, the operator $L_0$ can be expressed in terms of the bosons
$\phi^i(z)$,
\begin{align}
  L_0 &= \frac{1}{2} \sum_{i=1}^8 (F^i)^2 + \sum_{i=1}^8 N^i_{osc},
  \label{eq:akmmass}
\end{align}
where $N_{osc}^i$ is the number operator for the boson
$\phi^i(z)$. Thus, the weights of states including their
multiplicities at a given conformal weight can be obtained by solving
(\ref{eq:akmmass}). In the untwisted sector, the zero--modes of the
currents $\Omega^{IJ}(z)$ are the generators of
$\SO(16)$--transformations. They have conformal weight $0$ and
consequently the weights obtained by solving (\ref{eq:akmmass}) have
to form complete sets of weights of representations of $\SO(16)$.

The momenta $(p^i)$ are constraint to take values in
$\Lambda_{\SO(16)}$ and its spinor coset $\lambda_s +
\Lambda_{\SO(16)}$. The spinor weight $\lambda_s$ satisfies $2\lambda_s
\in \Lambda_{\SO(16)}$. Hence, the union of $\Lambda_{\SO(16)}$ and the
spinor coset is a lattice,
\begin{align}
  \Lambda_{\E_8} &= \Lambda_{\SO(16)} \cup \left(\lambda_s +
    \Lambda_{\SO(16)}\right),
\end{align}
which turns out to be isomorphic to the root lattice of the group
$\E_8$. Moreover, all roots $\alpha \in \Lambda_{\E_8}$ have
length $\alpha^2 = 2$. This implies that the vertex operator
$V_{\alpha}(z)$ is a current for every root of $\E_8$. By considering
the bosonic realization of Yang--Mills symmetry, it is possible to
show that the currents $E_\alpha(z) = c_\alpha V_\alpha(z)$ and
$H^i(z)$ satisfy an AKM algebra for
$\E_8$ \cite{Goddard1986}. $c_\alpha$ is a \emph{Klein cocycle} ensuring
the correct commutation relations. Thus, the AKM symmetry is enhanced from
$\SO(16)$ to $\E_8$. The additional currents correspond to the spin
fields $S^\alpha(z)$ creating the ground states in the R--sector
from the NS-vacuum. They have conformal weight $1$ by equation
(\ref{eq:fermionicshift}).

The same analysis can be repeated for the antiholomorphic fermions, which
represent an $\SO(8)$--AKM symmetry. The fermions can be bosonized by
four bosons $\phi^i(z)$. In this case the GSO projection in
the NS--sector is changed. The NS--ground state is assigned negative
$G$ parity. It turns out, that the momenta $q^i$ of the bosons $\phi^i$ in
the NS--sector is given by,
\begin{align}
  q^i &= \xi^i + n^i,\;\;\sum_{i=1}^4 n^i \in 2\mathbb{Z} + 1.
\end{align}
The weight vector $\lambda_v = (1,0,0,0)$ is a highest weight vector
for the vector representation of $\SO(8)$. By the same arguments that
led to (\ref{eq:ymmomenta}), it is possible to deduce that
\begin{subequations}
  \begin{align}
    (q^i) &\in \xi + \lambda_v + \Lambda_{\SO(8)}\mbox{ in the NS--sector} \\
    (q^i) &\in \xi + \lambda_s + \Lambda_{\SO(8)}\mbox{ in the R--sector,}
  \end{align}
  \label{eq:fermionmomenta}
\end{subequations}
where $\Lambda_{\SO(8)}$ denotes the root lattice of $\SO(8)$. In the
untwisted sector, the spin fields $S^\alpha(z)$ have conformal weight
$\frac{1}{2}$ by (\ref{eq:fermionicshift}) and are not currents,
hence, there is no enhanced AKM symmetry.

\subsection{Characters and modular properties}
In the fermionic realization of Yang--Mills symmetry there are eight
independent complex fermions. It is possible to define a
different gradation to every complex fermion. The operator $G_t$
generating the gradation has therefore eight parameter $t^i$ and acts
on the fermionic modes $b^i_n$ and $\bar{b}_n^i$ by
\begin{subequations}
  \begin{align}
    G_t b^i_n G_t^{-1} &= \exp\left( - 2\pi\imath t^i \right) b^i_n \\
    G_t \bar{b}^i_n G_t^{-1} &= \exp\left( 2\pi\imath t^i \right) \bar{b}^i_n.
  \end{align}
  \label{eq:gengrad}
\end{subequations}
The vacuum is again assigned $G_t$-charge $0$. Define the GSO
projected characters $\mathcal{Z}_{\xi,t}(\tau)$ by
\begin{align}
  \mathcal{Z}_{\xi,t}(\tau) &= \mbox{Tr} \frac{1+G}{2} G_t q^{L_0-8/24},
\end{align}
where $G$ is the $G$-parity operator. The character
$\mathcal{Z}_{\xi,t}(\tau)$ counts the number of states weighted by
their $G_t$-charge in the GSO-projected state space. It receives
contributions from both the NS-sector and the R-sector.

The complex fermions in the NS-sector are described by the twist
parameter $s^i = \frac{1}{2} + \xi^i$ according to equation
(\ref{eq:twistns}). The state space is a tensor product of the eight
complex fermion state spaces. The action of the operators $G_t$ on the
$i$-th state space is exactly the one considered in section
(\ref{sec:fermionorbifold}). The effect of the $G$-parity operator can
be included to the operator $G_t$ by defining $G_{t'}$ with $t'^i =
t^i + \frac{1}{2}$. Then it holds true that $G_{t'} = GG_t$. Using the
tensor product structure of the state space, one can show that the
contribution of the NS-sector to 
$\mathcal{Z}_{\xi,t}(\tau)$ is,
\begin{align}
  \mathcal{Z}_{NS}^{(\xi,t)}(\tau) &= \mbox{Tr}_{NS}
  \left(  \frac{1+G}{2}G_t q^{L_0 - 8/24} \right)
  = \notag \\ &= \frac{1}{2} \left( \prod_{i=1}^8\chi_{\xi^i+\frac{1}{2},t^i}(\tau) +
     \prod_{i=1}^8\chi_{\xi^i+\frac{1}{2},t^i+\frac{1}{2}}(\tau) \right),
\end{align}
where $\chi_{s,t}(\tau)$ is the contribution from one twisted complex
fermion. The functions $\chi_{s,t}(\tau)$ can be expressed in terms of
Jacobi theta functions and read, using the conventions of \cite{Mumford},
\begin{align}
  \chi_{s,t}(\tau) &= e^{-2\pi\imath \left(s-\frac{1}{2}\right) t} \frac{
      \vartheta_{s-\frac{1}{2},t}\left( 0, \tau\right)}{\eta(\tau)}.
\end{align}
In the R-sector the twist parameters are  $s^i = \xi^i$ according to
equation (\ref{eq:twistr}). By a similar calculation as
in the NS-sector, the contribution  $\mathcal{Z}_{R}^{(\xi,t)}(\tau)$
of the R-sector can be obtained and reads as
\begin{align}
  \mathcal{Z}_{R}^{(\xi,t)}(\tau) &= \mbox{Tr}_{R}
    \left( \frac{1+G}{2}G_t q^{L_0 - 8/24} \right)
  = \notag \\ &= \frac{1}{2} \left( \prod_{i=1}^8 \chi_{\xi^i,t^i}(\tau) +
    \prod_{i=1}^8 \chi_{\xi^i,t^i+\frac{1}{2}}(\tau) \right).
\end{align}
The full character $\mathcal{Z}_{\xi,t}(\tau)$ is the given by
\begin{align}
  \mathcal{Z}_{\xi,t}(\tau) &=  \mathcal{Z}_{NS}^{(\xi,t)}(\tau) +
  \mathcal{Z}_{R}^{(\xi,t)}(\tau).
\end{align}
Consider now the behaviour of the contributions
$\mathcal{Z}_{NS/R}^{(\xi,t)}(\tau)$ under $T$-modular
transformations. Using the modular properties of the Jacobi theta
function one can show that
\begin{subequations}
\begin{align}
  \mathcal{Z}_{NS}^{(\xi,t)}(\tau) &\overset{T}{\rightarrow}
  e^{-\frac{8\imath\pi}{12}}e^{\pi\imath \xi^2}
  \mathcal{Z}_{NS}^{(\xi,t+\xi)}(\tau) \\
  \mathcal{Z}_{R}^{(\xi,t)}(\tau) &\overset{T}{\rightarrow}
  e^{-\frac{8\imath\pi}{12}} e^{\pi\imath \left(\xi^2 - \sum_i \xi^i
    \right) }\mathcal{Z}_{R}^{(\xi,t+\xi)}(\tau).
\end{align}
\end{subequations}
The contributions do not transform homogeneously under $T$-modular
transformations and the characters $\mathcal{Z}_{\xi,t}(\tau)$ are not
mapped to themselves, unless
\begin{align}
  \sum_i \xi^i &\equiv 0\mbox{ mod }2.
\end{align}
Similar considerations for $S$-modular transformations require in
addition that the gradation parameters $t^i$ have to obey
\begin{align}
  \sum_i t^i &\equiv 0\mbox{ mod }2.
\end{align}
If the twist vector $\xi$ and the gradation parameter $t$ both satisfy
their restrictions, the set of characters is closed under modular
transformations and it can be shown
that
\begin{subequations}
\begin{align}
  \mathcal{Z}_{\xi,t}(\tau) &\overset{T}{\rightarrow}
  e^{-\frac{8\imath\pi}{12}}e^{\pi\imath \xi^2}
  \mathcal{Z}_{\xi,t+\xi}(\tau) \\
  \mathcal{Z}_{\xi,t}(\tau) &\overset{S}{\rightarrow}
  e^{-2 \pi\imath \xi \cdot t }\mathcal{Z}_{t,-\xi}(\tau).
\end{align}
\label{eq:fermionakmchars}
\end{subequations}
The calculations are the same for the anti-holomorphic fermions
representing $\SO(8)$-Lorentz symmetry. The central charge of this CFT
is $c=4$ and the phase factor under $T$-modular transformations is
changed. For completeness, their modular transformation properties are
given,
\begin{subequations}
\begin{align}
  \mathcal{Z}_{\xi,t}(\tau) &\overset{T}{\rightarrow}
  e^{-\frac{4\imath\pi}{12}}e^{\pi\imath \xi^2}
  \mathcal{Z}_{\xi,t+\xi}(\tau) \\
  \mathcal{Z}_{\xi,t}(\tau) &\overset{S}{\rightarrow}
  e^{-2 \pi\imath \xi \cdot t }\mathcal{Z}_{t,-\xi}(\tau).
\end{align}
\label{eq:fermionso8chars}
\end{subequations}

\section{Tables and pictures}
\begin{sidewaystable}[hbtp]
\begin{center}
  \begin{tabular}{c||c|c|c|c|}
    &{\sl $N=1$ SUSY multiplet} & {\sl $\E_6\times \U(1)$ charges} &
    {\sl Degeneracy} & {\sl Interpretation} \\[.3em]
    \hline
    \hline
    \multirow{8}{*}{$[\iota]$} &
    Gravity $(0,0)$ & $\mathbf{1}_0$ & 1 & SUGRA multiplet \\
    &Chiral $(1,0)$ & $\mathbf{1}_0$ & 1 & Dilaton/Dilatino \\
    &Chiral $(1,0)$  & $\mathbf{1}_0$ & 2 & K\"ahler moduli \\
    &Chiral $(1,0)$  & $\mathbf{1}_0$ & 2 & Complex structure moduli \\[.3em]
    \cline{2-5}
    &Vector $(0,0)$ &  $\mathbf{78}_0 \oplus \mathbf{1}_0$ & 1 & YM
    gauge bosons \\
    &Vector $(0,0)$ &  $248 \times \mathbf{1}_0$ & 1 & YM
    gauge bosons \\
    &Chiral $(-1,0)$ & $\mathbf{27}_2 \oplus \mathbf{\overline{27}}_{-2} \oplus
   \mathbf{1}_0$ & 1 & Charged matter \\
   &Chiral $(1,0)$ & $\mathbf{27}_{-1} \oplus \mathbf{\overline{27}}_{1} $ & 1 &
   Charged matter  \\
   &Chiral $(1,0)$ & $\mathbf{1}_0 \oplus \mathbf{1}_{3} \oplus
   \mathbf{1}_{-3}$ & 1 &
   Charged matter  \\[.3em]
   \hline \hline

   \multirow{5}{*}{$[\tau]$} &
    Chiral $(2,1)$ & \multirow{4}{*}{$\mathbf{1}_{3/2} \oplus \mathbf{1}_{-3/2}$} & \multirow{4}{*}{4} &
    \multirow{4}{*}{Charged matter} \\ 
    &Chiral $(2,1)$ &  & & \\ 
    &Chiral $(0,1)$ &  & & \\ 
    &Chiral $(0,1)$ &  & & \\[.3em]
    \cline{2-5}
    &Chiral $(0,1)$ & $\mathbf{1}_{3/2} \oplus \mathbf{1}_{-3/2} \oplus
    \mathbf{27}_{1/2} \oplus \overline{\mathbf{27}}_{-1/2}$ & 4 &
    Charged matter  \\[.3em]
   \hline \hline

   \multirow{2}{*}{$[\sigma]$} &
    Chiral $(1,0)$ & \multirow{1}{*}{$7 \times \mathbf{1}_{0}$} & \multirow{1}{*}{9} &
    \multirow{1}{*}{Charged matter} \\[.3em]
    \cline{2-5}
    & Chiral $(1,0)$ & $\mathbf{1}_{3} \oplus \mathbf{1}_{-3} \oplus
    \mathbf{27}_{-1} \oplus \overline{\mathbf{27}}_{1}$ & 9 &
    Charged matter  \\[.3em]
   \hline 
  \end{tabular}
\end{center}
\caption{The full physical, massless spectrum of the $S_3$ model. It
  shows the different types of massless particles together with a
  possible interpretation and quantum numbers, like geometric
  transformation properties and representations under the $\E_6\times
  \U(1)\times \E_8$ low--energy Yang--Mills gauge group. However, all
  particles are uncharged under $\E_8$ and $\E_8$ charges are not shown
  explicitly. The given degeneracy is due to the geometry (fixed tori,
  etc.). The additional quantum numbers $(r,c)$ are the charges under
  the $R$ symmetry $\mathbb{Z}_{4}^R$ and the $\mathbb{Z}_2$ parity
  introduced in section \ref{sec:fermionsector}. Only the $R$ charge
  of the state with highest helicity is given.}
\label{tab:s3modelspectrum}
\end{sidewaystable}

\begin{sidewaystable}[hbtp]
  
  \begin{center}
  \begin{tabular}{c||c|c|c|c|c|c|c|}
    {\sl Sector} & 
    {\sl Conformal weight} & {\sl $4d$ Type} & {\sl Centralizer
      irrep.} & {\sl Degeneracy} & {\sl Spin sector} & {\sl
      $D$ charge} & {\sl $R$ and $\mathbb{Z}_2$ charge} \\[.3em]
    \hline\hline
    \multirow{6}{*}{$\iota$} &
    \multirow{6}{*}{$1/2$} & vector & $r_{\mathrm{inv}}$ & 1 & NS & $(0,0)$ & $(0,0)$ \\
          && scalar & $r_{\mathrm{alt}}$ & 2 & NS  & $(0, \mp 2)$ & $ (2, 0)$ \\
          && scalar & $r_{\mathrm{def}}$ & 2 & NS  & $(\pm 2, 0)$ & $ (0,0)$ \\
    && Weyl & $r_{\mathrm{inv}}$ & 1 & R & $(-2,-1)$ & $ (-1,0)$ \\
          && Weyl & $r_{\mathrm{alt}}$ & 1 & R & $(2, - 1)$ & $ (-1,0)$ \\
          && Weyl & $r_{\mathrm{def}}$ & 1 & R & $(0,1)$  & $ (1,0)$ \\[.3em]
          \hline\hline

    \multirow{2}{*}{$\tau$} &
    \multirow{2}{*}{$1/4$} & scalar & $r_{+}$ & 2 & NS & $(\pm 1, \mp
    1)$ & $ (\mp 1, 1 )$ \\
    && Weyl & $r_{+}$ & 1 & R & $(- 1, 0)$ & $(0,1)$ \\[.3em]
          \hline\hline

     \multirow{2}{*}{$\sigma$} &
     \multirow{2}{*}{$5/18$} & scalar & $r_{0}$ & 2 & NS & $(\pm 2,
     0)$ & $ (0,0)$ \\
    && Weyl & $r_{0}$ & 1 & R & $(0,1)$ & $(1,0)$ \\[.3em]
          \hline
  \end{tabular}
  \end{center}

  \caption{String states with conformal weight $\leq 1/2$ in the
    fermion sector of the $S_3$ model. It is shown the conformal
    weight, the transformation properties in Minkowski space, the
    degeneracy and the representation under the centralizer group and the
    $D$ charges of discrete symmetry
    $\mathbb{Z}_{8}\times\mathbb{Z}_{12}$. Only the $D$ charges of
    the highest helicity states are shown.} 
  \label{tab:fermionsector}
\end{sidewaystable}

\begin{sidewaystable}[hbtp]
  \begin{adjustwidth}{-3cm}{-3cm}
  \begin{center}
  \begin{tabular}{c||c|c|c||c|c}
    {\sl Sector} & 
    {\sl Conformal weight} & {\sl $\E_6 \times \U(1)$ quantum numbers} &
    {\sl Centralizer irrep.} & \multicolumn{2}{c}{\sl $\SO(10)\times
      \U(1)\times \U(1)$ quantum numbers} \\[.3em]
    & & & & {\sl NS-sector} & {\sl R-sector} \\[.3em]
    \hline\hline
    \multirow{4}{*}{$\iota$} &
    $0$ & $\mathbf{1}_0$ & $r_{\mathrm{inv}}$ & $\mathbf{1}_{0,0}$ &  \\[.3em]
          \cline{2-6}
    &\multirow{3}{*}{$1$} & $\mathbf{78}_0$  &
    \multirow{2}{*}{$r_{\mathrm{inv}}$} & $\mathbf{45}_{0,0} \oplus \mathbf{1}_{0,0}$ &
    $\mathbf{16}_{-1,1/2} \oplus \overline{\mathbf{16}}_{1,-1/2}$\\ 
   & & $\mathbf{1}_0$ & & $\mathbf{1}_{0,0}$ & \\[.3em]
    \cline{3-6}
   && $\mathbf{27}_2 \oplus \mathbf{\overline{27}}_{-2}$ &
   \multirow{2}{*}{$r_{\mathrm{alt}}$} &
   $\mathbf{10}_{0,1} \oplus \mathbf{1}_{2,0} \oplus \mbox{c.c.}$ & 
   $\mathbf{16}_{1,1/2} \oplus \mbox{c.c.}$ \\
  && $\mathbf{1}_0$ & & $\mathbf{1}_{0,0}$ &  \\[.3em]
   \cline{3-6}
  && $\mathbf{27}_{-1} \oplus \mathbf{\overline{27}}_{1} $
   & \multirow{3}{*}{$r_{\mathrm{def}}$}  & $\mathbf{10}_{-1,0} \oplus
   \mathbf{1}_{1,-1} \oplus \mbox{c.c.}$ &
   $\mathbf{16}_{0,-1/2} \oplus \mbox{c.c.}$ \\
  && $\mathbf{1}_{3} \oplus \mathbf{1}_{-3}$
   & & $\mathbf{1}_{1,1} \oplus \mbox{c.c.}$ & \\
  && $\mathbf{1}_0$
   & & $\mathbf{1}_{0,0}$ & \\[.3em]
          \hline\hline

          \multirow{2}{*}{$\tau$} &
    $1/4$ & $\mathbf{1}_{3/2} \oplus \mathbf{1}_{-3/2}$ & $r_{-}$ &
    $\mathbf{1}_{1/2,1/2} \oplus \mbox{c.c.}$ &  \\[.3em]
          \cline{2-6}
    & \multirow{2}{*}{$3/4$} & $\mathbf{27}_{1/2} \oplus \mathbf{\overline{27}}_{-1/2} $
   & \multirow{2}{*}{$r_{+}$} & $ \mathbf{10}_{-1/2,1/2} \oplus
   \mathbf{1}_{3/2,-1/2} \oplus
   \mbox{c.c.}$ & $ \mathbf{16}_{1/2,0} \oplus \mbox{c.c.}$ \\
   &  & $\mathbf{1}_{3/2} \oplus \mathbf{1}_{-3/2}$
   & &$ \mathbf{1}_{1/2,1/2} \oplus
   \mbox{c.c.}$ &   \\[.3em]
          \hline\hline

        \multirow{3}{*}{$\sigma$} &
    $1/9$ & $\mathbf{1}_{0}$ & $r_{-}$ & $\mathbf{1}_{0,0}$& \\[.3em]
          \cline{2-6}
    &$4/9$ & $\mathbf{1}_{0}$ & $r_{+}$ & $\mathbf{1}_{0,0}$& \\[.3em]
          \cline{2-6}
    &\multirow{2}{*}{$7/9$} & $\mathbf{27}_{-1} \oplus
    \mathbf{\overline{27}}_{1}$ & \multirow{2}{*}{$r_{0}$} &
    $\mathbf{10}_{-1,0} \oplus \mathbf{1}_{1,-1} \oplus\mbox{c.c.}  $
    & $\mathbf{16}_{0,-1/2} \oplus\mbox{c.c.} $ \\ 
   & & $ \mathbf{1}_{3} \oplus
    \mathbf{1}_{-3}$ & & $\mathbf{1}_{1,1}\oplus\mbox{c.c.}$ &   \\[.3em] 
          \hline

  \end{tabular}
  \end{center}
\end{adjustwidth}
\caption{String states with conformal weight $\leq 1$ in the
    Yang--Mills sector of the $S_3$-model. The conformal
    weight, the representation under the Yang--Mills gauge group as
    well as the representation under the centralizer group are shown. On the
    right hand side the decomposition into irreducible representations
    of $\mathfrak{g}_{0}$ is shown.}
  \label{tab:ymsector}
\end{sidewaystable}


\begin{thebibliography}{[20]}
\bibitem{HeteroticString1} Gross, D., et al., \emph{Heterotic String
    Theory. 1. The Free Heterotic String}, Nucl.\ Phys.\ B256, 253,
  1985
\bibitem{HeteroticString2}
  Gross, D., et al., \emph{The Heterotic String}, Phys.\ Rev.\
  Lett. 54, 502, 1985
\bibitem{HeteroticString3}
  Gross, D., et al., \emph{Heterotic String
    Theory. 2. The Interacting Heterotic String}, Nucl.\ Phys.\ B267, 75,
  1986
\bibitem{Raby} Raby, S., \emph{String Model Building}, AIP Conf.\
  Proc.\ 1200, 235, 2010
\bibitem{Antoniadis1} Antoniadis, I., et al., \emph{GUT model building
  with fermionic four-dimensional strings}, Phys.\ Lett.\ B205, 1988
\bibitem{Aldazabal} Aldazabal, G., et al., \emph{Building GUTs from
    strings}, Nucl.\ Phys.\ B465, 34, 1996
\bibitem{StringOrbifolds1} Dixon, L., et al., \emph{Strings on
    orbifolds}, Nucl.\ Phys.\ B261, 678, 1985
\bibitem{StringOrbifolds2} Dixon, L., et al., \emph{Strings on
    orbifolds (ii)}, Nucl.\ Phys.\ B274, 285, 1986
\bibitem{FischerOrb} Fischer, M., et al., \emph{Classification of
    symmetric toroidal orbifolds}, JHEP 1301 (2013) 084, [arXiv:hep-th/1209.3906]
\bibitem{Lebedev1} Lebedev, O., et al., \emph{A mini landscape of
    exact MSSM spectra in heterotic orbifolds}, Phys.\ Lett.\ B645,
  88, 2007
\bibitem{Lebedev2} Lebedev, O., et al., \emph{Heterotic road to the
    MSSM with R parity}, Phys.\ Rev.\ D77, 046013, 2008
\bibitem{Blaszczyk1} Blaszczyk, M., et al., \emph{A
    $\mathbb{Z}_2\times\mathbb{Z}_2$ standard model}, Phys.\ Lett.\
  B683, 340, 2010, [arXiv:hep-th/0911.4905]
\bibitem{Kappl1} Kappl, R., et al., \emph{String derived MSSM vacua
    with residual R symmetries}, Nucl.\ Phys.\ B847, 325, 2011,
  [arXiv:hep-th/1012.4574]
\bibitem{Lebedev3} Lebedev, O., et al., \emph{Heterotic mini land-space
  (II): completing the search for MSSM vacua in $\mathbb{Z}_6$
  orbifolds}, Phys.\ Lett.\ B668, 331, 2008, [arXiv:hep-th/0807.4384]
\bibitem{Ibanez1} Ib\'a\~nez, L. E., et al., \emph{Heterotic strings
    in symmetric and asymmetric backgrounds}, Nucl.\ Phys.\ B301, 157, 1988
\bibitem{Ibanez2} Ib\'a\~nez, L. E., et al., \emph{Reducing the rank
    of the gauge group in orbifold compactifications of the heterotic
    string}, Phys.\ Lett.\ B192, 332, 1987
\bibitem{Dreiner} Dreiner, H., et al., \emph{String model building in
    the free fermionic formulation}, Nucl.\ Phys.\ B320, 401, 1989
\bibitem{Antoniadis} Antoniadis, I. and Bachas, C., \emph{4d fermionic
  superstring with arbitrary twists}, Nucl.\ Phys.\ B298, 586, 1988
\bibitem{Kakushadze} Kakushadze, Z., et al., \emph{Asymmetric
    non-Abelian orbifolds and model building}, Phys.\ Rev.\ D54, 7545,
  1986, [arXiv:hep-th/9607137]
\bibitem{WittenNA} Witten, E., \emph{Non-Abelian bosonization in two
    dimensions}, Commun.\ Math.\ Phys.\ 92, 455, 1984
\bibitem{Aoki} Aoki, K., et al., \emph{On the construction of
    asymmetric orbifold models}, Nucl.\ Phys.\ B695, 132, 2004,
  [arXiv:hep-th/0402134]
\bibitem{Feng} Feng, B. et al., \emph{Discrete torsion, non-Abelian
    orbifolds and the Schur multiplier}, JHEP 0101, 033, 2001
\bibitem{KonopkaDiplomaThesis} Konopka, S., \emph{Non-Abelian orbifold
    compactifications of the heterotic string}, Diploma thesis,
  Technische Universit\"at M\"unchen, December 2011,
  \texttt{\small
    http://einrichtungen.ph.tum.de/T30e/research/theses/KonopkaDiplomarbeit.pdf}
\bibitem{Ginsparg} Ginsparg, P., \emph{Curiosities at $c=1$}, Nucl.\
  Phys.\ B295, 153, 1988
\bibitem{Dijkgraaf} Dijkgraaf, R., et al., \emph{The operator algebra
    of orbifold models}, Commun.\ Math.\ Phys. 123,485, 1989
\bibitem{RatzStringy} Ratz, M., et al., \emph{Stringy surprises},
  Yukawa International Seminar Symposium, 2009,
  [arXiv:hep-th/1003.0549]
\bibitem{Lee} Lee, H. M., et al., \emph{A unique symmetry for the
    MSSM}, Phys.\ Lett.\ B694, 491, 2011
\bibitem{GAP} The GAP group, \emph{GAP - Groups, Algorithms and
    Programming, Version 4.4.12}, 2008,
  \texttt{\small http://www.gap-system.org/}
\bibitem{Fulton} Fulton, W. and Harris, J., \emph{Representation
    Theory}, volume 129 of \emph{Graduate Texts in Mathematics},
  Springer--Verlag, 1991
\bibitem{Milgram} Adem, A. and Milgram, R. J., \emph{Cohomology of
    finite groups}, Springer--Verlag, 2004
\bibitem{Vafa} Vafa, C., \emph{Modular invariance and discrete
    torsion on orbifolds}, Nucl.\ Phys.\ B273, 592, 1986
\bibitem{MirageTorsion} Pl\"oger, F., et al., \emph{Mirage Torsion},
  JHEP 0704, 063, 2007
\bibitem{Fujikawa} Fujikawa, K., \emph{Comment on chiral and conformal
    anomalies}, Phys.\ Rev.\ Lett. 44, 1733, 1980
\bibitem{Green} Green, M. and Schwarz, J., \emph{Anomaly cancellation
    in supersymmetric $D=10$ gauge theory and superstring theory},
  Phys.\ Lett.\ B149, 117, 1984
\bibitem{Tosa} Tosa, Y., \emph{Generalized Green--Schwarz anomaly
    cancellation mechanism}, Phys.\ Rev.\ D39, 1648, 1989
\bibitem{Chen} Chen, M.--C., et al., \emph{The $\mu$ term and neutrino
    masses}, [arXiv:hep-ph/1206.5375]
\bibitem{Schwarz:1995zw} 
  Schwarz, J.~H.~,\emph{Anomaly - free supersymmetric models in six-dimensions},
  Phys.\ Lett.\ B {\bf 371}, 223 (1996),
  [hep-th/9512053]
\bibitem{Witten:1995gx} Witten, E., \emph{Small instantons in string theory},
  Nucl.\ Phys.\ B {\bf 460} (1996) 541 [hep-th/9511030]
\bibitem{Aldazabal:1997wi} 
  Aldazabal, G., et.~al.,
  \emph{Nonperturbative heterotic D = 6, D = 4, N=1 orbifold vacua},
  Nucl.\ Phys.\ B {\bf 519}, 239 (1998)
  [hep-th/9706158]
\bibitem{Fischer:2013qza} 
  M.~Fischer, et.~.al.,\emph{Heterotic non-Abelian orbifolds},
  arXiv:1304.7742 [hep-th]
\bibitem{Goddard1986} Goddard, P. and Olive, D., \emph{Ka\v{c}--Moody
    and Virasoro algebras in relation to quantum physics}, Int.\ J.\
  Mod.\ Phys. 1, 303, 1986
\bibitem{Mumford} Mumford, D., \emph{Tata lectures on theta I--III},
  Birkh\"auser Boston, 1992
\end{thebibliography}
\end{document}